\documentclass[10pt,journal]{IEEEtran}

\usepackage{cite}

\ifCLASSINFOpdf
  \usepackage[pdftex]{graphicx}
\else
  \usepackage[dvips]{graphicx}
\fi

\usepackage[cmex10]{amsmath}
\usepackage{amsfonts}

\usepackage{subfloat}
\usepackage{subfig}
\usepackage{stfloats}
\usepackage{multirow}
\usepackage{color}
\usepackage{url}
\usepackage[font=small]{caption}
\usepackage{array}

\providecommand{\ie}{\emph{i.e.,} }
\providecommand{\eg}{\emph{e.g.,} }
\providecommand{\parab}[1]{\noindent\textbf{#1}}
\providecommand{\para}[1]{\noindent\textit{#1}}

\newcolumntype{C}[1]{>{\centering\let\newline\\\arraybackslash\hspace{0pt}}m{#1}}

\begin{document}

\title{Mining MOOC Clickstreams: On the Relationship Between Learner Behavior and Performance}

\author{Christopher~G.~Brinton,~\IEEEmembership{Student Member,~IEEE,}
	Swapna~Buccapatnam,~\IEEEmembership{Member,~IEEE,}
	Mung~Chiang,~\IEEEmembership{Fellow,~IEEE,}
	H.~Vincent~Poor,~\IEEEmembership{Fellow,~IEEE}
\thanks{The authors are with the Department
of Electrical Engineering at Princeton University, Princeton, NJ 08544. Email: \{cbrinton, swapnab, chiangm, poor\}@princeton.edu.}}

\maketitle

\begin{abstract}
We study student behavior and performance in two Massive Open Online Courses (MOOCs). In doing so, we present two frameworks by which video-watching clickstreams can be represented: one based on the sequence of events created, and another on the sequence of positions visited. With the event-based framework, we extract recurring subsequences of student behavior, which contain fundamental characteristics such as reflecting (\ie repeatedly playing and pausing) and revising (\ie plays and skip backs). We find that some of these behaviors are significantly associated with whether a user will be Correct on First Attempt (CFA) or not in answering quiz questions, and in ways that are not necessarily intuitive. Then, with the position-based framework, we devise models of performance based naturally on user behavior. In evaluating these models through CFA prediction, we find that three of them can substantially improve prediction quality in terms of accuracy and F1, which underlines the ability to relate behavior to performance. Since our prediction considers videos individually, these benefits also suggest that our models are useful in situations where there is limited training data, \eg for early detection or in short courses.
\end{abstract}


\IEEEpeerreviewmaketitle

\section{Introduction}
\label{sec:introduction}
\IEEEPARstart{O}{ver} the past decade, technology advances have been influencing the ways we can learn. One of the prominent innovations has been the Massive Open Online Course (MOOC).  MOOC providers such as Coursera, edX, and Udacity have offered courses reaching out to tens, and even hundreds of thousands of students within single sessions \cite{brinton2014social}.

One salient feature of MOOC is its high attrition rates, with typically less than 10\% of students initially enrolled in a course seeing it to completion. These low completion rates, attributed to factors such as small teacher-to-student ratios, the asynchronous nature of interaction, and diverse demographics, have made MOOC the subject of controversy as the future of higher education is explored \cite{brinton2015individualization}. This has in turn ignited a growing body of research interest in understanding why these dropoff rates occur \cite{brinton2014learning,anderson2014engaging}, and in designing mechanisms to improve the quality of learning on MOOCs, such as: through early detections of students with low performance \cite{brinton2015prediction} or high dropoff likelihoods \cite{yang2013turn,sinha2014your}, through recommendations for discussion participation \cite{brinton2014learning} or for certain allocations of peer grading \cite{piech2013tuned}, and through automated individualization \cite{brinton2015individualization}.

A standard MOOC will contain three different learning modes for students: video lectures, assessments (\eg in-video quizzes, homework assignments, and exams), and social networking (usually through discussion forums) \cite{brinton2015individualization}. Most platforms track student interaction with these different forms of learning, with backends designed to collect measurements as a student navigates through the course. For video content, these measurements include clickstream events, with a click event being generated and stored each time a learner interacts with a video, specifying the particular action (\eg pause, skip), position, and time at which it occurred. For assessments, the specific responses to individual questions are tracked. For the discussion forums, the sequence of posts and comments are stored. This type of big data has been the focus of a number of recent studies in machine learning and data mining on understanding how MOOC users learn \cite{kim2014understanding,sinha2014your,brinton2014learning}.

\begin{figure*}
\centering
\begin{tabular}{c|cccc|cccc}
\multirow{2}{*}{Dataset} & \multirow{2}{*}{Lectures} & Lecture & Video Length (min) & \multirow{2}{*}{Quizzes} & \multirow{2}{*}{Users} & Clickstream & User-Video & CFA Score \\
	& & Videos & avg. (s.d.) & &  & Events & Pairs & avg. (s.d.) \\ \hline
`FMB' & 20 & 92 & 16.9 (5.96) & 92 & 3770 & 314,632 & 26,250 & 0.663 (0.473) \\
`NI' & 6 & 115 & 5.44 (2.17) & 69 & 2680 & 416,214 & 36,464 & 0.750 (0.433) \\
\end{tabular}
\caption{Basic information on the two datatsets. The values in the right column are the final numbers after data filtering.}
\label{fig:basicInfo}
\vspace{-0.15in}
\end{figure*}

\parab{Motivation and objectives}. What remains understudied, however, is the \textit{relationship} between these learning modes. In particular, is it possible to associate a student's \textit{behavior} with his/her \textit{performance} in a MOOC? Developing such an understanding would have implications not only to theories about how humans process information, but also to systems for improving low completion rates. For example, systems for individualized content delivery have largely been driven by algorithms that model users solely based on their assessment performance. This tends to be a sparse source of information about users in MOOCs, since many users complete few assessments \cite{brinton2015prediction}. Uncovering relationships between behavior and performance would allow individualization algorithms to be augmented with behavioral signals to determine the most suitable path of learning for each student to take, as suggested in \cite{brinton2015individualization}. These relationships could also be provided to course instructors directly, in the form of extended learning analytics \cite{stephens2014monitoring}, to give instructors insight into which parts of their content contribute to more effective learning outcomes in their courses.

Our work is motivated by this fundamental question of if, and how, it is possible to relate behavior to performance. In our investigation, we focus on the video-watching behavior of MOOC students, where users spend the majority of their time learning \cite{kim2014understanding}. These videos are typically equipped with quiz questions, which serve as immediate feedback of the knowledge a student gained from the content in the video. In relating behavior to performance, then, we can consider (i) the clickstreams generated by a user in watching the video associated with a particular quiz (\ie the behavioral aspect), and (ii) whether the user was Correct on First Attempt (CFA) or not (non-CFA) in answering the given quiz question (\ie the performance aspect).

In our investigation, we formalize different ways that video-watching clickstreams can be represented as sequences, and apply the frameworks we develop to meet two objectives:

\para{Objective 1 (O1)}: To identify recurring behaviors of learners, such as revising content or skipping forward repeatedly.

\para{Objective 2 (O2)}: To assess the impact of behavior on performance, \ie how patterns identified in O1 and specific positions visited in each video are signals of effective learning.

Previous work on studying the video-watching clickstreams of students \cite{sinha2014your} has focused on the sequence of \textit{events} (\eg pause, skip forward) generated. In studying O1\&2, we identify two additional factors that are important to capture: the \textit{positions} in the video that a student visited, and the \textit{duration / length} of time between the events and positions. These form the basis of our clickstream representation frameworks.

In our investigation, we employ two datasets coming from two different MOOCs we have instructed on Coursera. After filtering (described in Sec. \ref{sec:datasets}), these datasets contain 315K and 416K clickstream events corresponding to 26K and 36K first-attempt quiz submissions by students. With these datasets, our study is specifically broken down into two components: behavioral motifs and behavior-based prediction, as follows.

\parab{Behavioral motifs}. We first develop an event-based framework to represent clickstreams (Sec. \ref{sec:datasets}), which captures event types and their lengths. Leveraging this framework, we are able to identify video-watching \textit{motifs}, \ie sub-sequences of student behavior that occur significantly often, in our two datasets. These motifs by themselves are informative of recurring behaviors for O1 (Sec. \ref{sec:motifs}), and additionally, we are able to identify a significant difference in the presence of certain motifs between the CFA and non-CFA sequences for O2. For example, we find that a series of behaviors are indicative of students reflecting on material, and are significantly associated with the CFA sequences in one of our courses. As another example, we identify motifs that are consistent with rapid-paced skimming through the material, and reveal that these are discriminatory in favor of non-CFA in both of our courses.

For these motifs, the identified association with CFA or non-CFA (when one exists) is particularly helpful, because for many of them, either case is conceivable. For one, skimming could intuitively be a sign of a student either correctly or incorrectly perceiving familiarity with the material; our results indicate the latter is more likely. Also, we find that incorporating the lengths in addition to the events is important to these findings, because extracting motifs from sequences of events alone does not reveal these insights.

\parab{Behavior-based prediction}. In investigating O2, we will also develop models for knowledge gained based on the clicks that a student makes in a video. The quality of such a model can be evaluated by considering its ability to generalize to incoming samples through prediction. The higher the quality, the stronger the association between behavior and performance.

To this end, we will also study student performance prediction (specifically, CFA prediction) for MOOC. Enhancing CFA prediction is an important area of research in its own right, because such methods can improve systems for early detection of \eg struggling/advanced students and easy/difficult material \cite{brinton2015prediction}. In seeking appropriate models for student performance, we find that while some behavioral patterns of the motifs are significantly associated with performance, their supports and the resulting success estimates are not sufficient to make large improvements in CFA prediction. As a result, we propose a second behavioral representation, which is based on the sequence of positions visited in a video (Sec. \ref{sec:positions}). In contrast to training over a long course duration as in \cite{brinton2015prediction,lan2014time}, we consider CFA prediction on a \textit{per-video} basis, in order to quantify the benefit obtained by the positions in each individual video.

We evaluate three different models based on our framework (Sec. \ref{sec:modelEval}), and find that they obtain substantial improvements in prediction when compared to a baseline that does not use click information. This underscores the ability to relate clicks to knowledge gained, \ie that behavior is related to performance, and shows that behavioral information is useful in situations where multiple videos are not be available, \eg in short courses or for detection early in a course. Further, since our algorithms are natural representations of student behavior (\eg sequences of positions visited), they can be used to guide student actions while watching a video in real time.

\parab{Summary of contribution}. Compared with other work (Sec. \ref{sec:related}), we make three main contributions in this paper:
\begin{enumerate}
\item We develop two new frameworks for representing student video-watching behavior as sequences.
\item We extract the recurring behavioral motifs of students watching videos using motif identification schemes, and associate these fundamental patterns with performance.
\item We demonstrate that video-watching behavior can be used to enhance student performance prediction on a per-video basis, \eg for earliest detection.
\end{enumerate}

\section{Datasets and Clickstreams}
\label{sec:datasets}
In this section, we describe our datasets, and present our first sequence specification based on events and lengths.

\subsection{Our Two MOOCs}
\label{ssec:ourCourses}
Our datasets come from two different courses that we have instructed on Coursera: \textit{Networks: Friends, Money, and Bytes} (`FMB') and \textit{Networks Illustrated: Principles Without Calculus} (`NI').\footnote{\url{www.coursera.org/course/{friendsmoneybytes,ni}}} Each of these courses teach networking topics, but `FMB' delves into the mathematical specifics behind the topics, whereas `NI' is meant as an introduction to the subject (see \cite{brinton2015individualization} for more details). We obtained two types of data from Coursera for each of the courses: (i) video-watching clickstreams, which log user interaction with the video player, and (ii) information on the in-video quiz submissions. We will describe the format of the video-watching clickstreams in detail in Sec. \ref{ssec:evSpec}, in developing a representation framework.

\parab{Course format}. The course formats are summarized in Fig. \ref{fig:basicInfo}. Each is made up of a series of lectures, which are in turn comprised of a set of videos. `FMB' is a longer course, with 20 lectures, whereas `NI' only has 6. `NI' had more, shorter-length videos, with a total of 115 videos and an average (avg.) length of 5.4 min per video, whereas `FMB' has less, longer-length videos, with 93 total and an avg. length of 16.9 min.

For each course, we included in-video quizzes at the end of the videos, to test a student's understanding of the material throughout the course. Each quiz is a multiple choice question, in radio-response format, with 4-5 possible answer choices. For `FMB', there was one question at the end of each video, whereas for `NI', each of the 69 questions was associated with anywhere from 1-4 videos. In mapping videos to quizzes, we will refer to ``video X'' as the contiguous set of videos occurring after question $X - 1$ and before question $X$.

\parab{User-Video Pairs}. We extract User-Video (UV) Pairs from the data, with two sets of information for video and quiz $X$:

\begin{figure}
\vspace{-0.1in}
\centering
\includegraphics[scale=0.29]{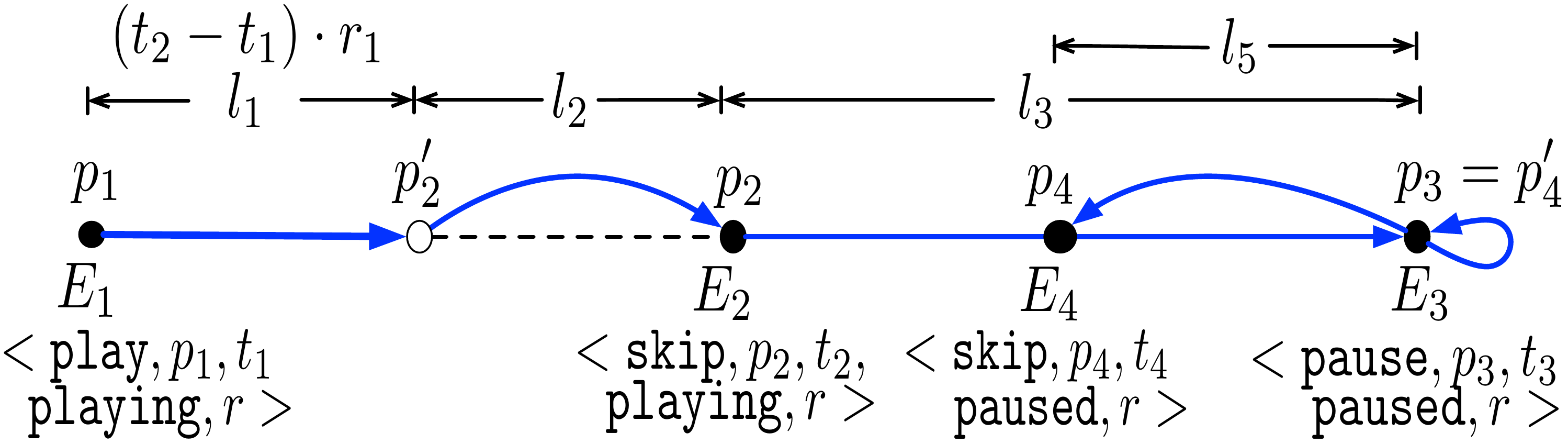}
\caption{Illustration of a sequence of clicks $E_1$ to $E_4$ on a video, where the horizontal axis denotes the video length. This will generate 5 events according to our first framework based on events and lengths. The length $l_j$ for the events that have this property (note that pauses do not have lengths) are depicted above the diagram.}
\label{fig:clickEvents}
\vspace{-0.2in}
\end{figure}

\para{(i) Video-watching trajectory}: The set of clickstream logs (events) for the user in video $X$.

\para{(ii) CFA result}: Whether the user was Correct on First Attempt (CFA) or not (non-CFA) for quiz $X$.

In total, there were 122.5K UV Pairs for `FMB', with 566K click events. For `NI', these numbers were 149K and 882K, respectively. After removing any UV Pair that had at least one null, stall or error contained in its video-watching trajectory, we were left with the numbers given in Fig. \ref{fig:basicInfo}. The avg. CFA score across the UV Pairs was $0.663$ for `FMB' (standard deviation (s.d.) $= 0.47$), and $0.750$ for `NI' (s.d. $= 0.43$).

\subsection{Processing Clickstream Events}
\label{sec:clickstream}
\subsubsection{Our nomenclature for events}
\label{ssec:evSpec}
A clickstream log is one of four types: {\tt play}, {\tt pause}, {\tt ratechange}, or {\tt skip}. Each time one of these events is fired, a data entry is recorded that specifies the user and video IDs, event type, playback position, playback speed, and UNIX timestamp for the event.

Formally, let $E_i$ denote the $i$th click event that occurs while a user is watching a video. We write $E_i = \langle e_i, p_i, t_i, s_i, r_i \rangle$, where $e_i$ is the type of the $i$th click, $p_i$ is the video position (in sec) right after $E_i$ is fired, $t_i$ is the UNIX time (in sec) at which $E_i$ was fired, $s_i$ is the state of the video player -- either {\tt playing} or {\tt paused} -- as a result of $E_i$, and $r_i$ is the playback rate (\ie speed) of the video player resulting from this event. The logs are sequenced chronologically for a UV Pair, \ie $t_1 < t_2 < \cdots$. Based on the $E_i$ for a UV Pair, we define the following events: 

\parab{Play} (Pl): A play event begins at the time when a click event $E_i$ is made for which the state $s_i$ is {\tt playing}, and lasts until the next click $E_{i+1}$. It occurs for a duration $d = t_{i+1} - t_i$ and has a length $l = p_{i+1} - p_i$.

\parab{Pause} (Pa): A pause event is defined in the same way as a play event, except it is for which the state $s_i$ is {\tt paused}, and does not have any length by definition.

\parab{Skip back} (Sb): A skip back (\ie rewind) event occurs when the type $e_i = {\tt skip}$ and $p'_i > p_i$, where $p'_i$ is the position of the video player immediately before the skip. If $s_{i-1} = {\tt playing}$, then $p'_i = p_{i-1} + (t_i - t_{i-1}) \cdot r_{i-1}$; if $s_{i-1} = {\tt paused}$, then $p'_i = p_{i-1}$. The length of the skip is $l = |p_i - p'_i|$, and there is no associated duration.

\parab{Skip forward} (Sf): A skip forward (\ie fast forward) event is defined as Sb, except it captures the case where $p_i > p'_i$.

\parab{Ratechange fast} (Rf): This occurs when $e_i = {\tt ratechange}$ and the rate $r_i > 1.0$.\footnote{On Coursera, the default player speed is 1.0, and users can vary this between 0.5 and 2.0, in increments of 0.25.} There is no duration or length.

\parab{Ratechange slow} (Rs): This occurs when $e_i = {\tt ratechange}$ and $r_i < 1$, again with no duration or length.

\parab{Ratechange default} (Rd): This is when $e_i = {\tt ratechange}$ and $r_i = 1$, \ie returning to the default.

With these, the sequence of events for a UV Pair becomes $\hat{e}_1, \hat{e}_2, ...$ for $\hat{e}_j \in \mathcal{E} = \{\mbox{Pl}, \mbox{Pa}, \mbox{Sb}, ...\}$, $|\mathcal{E}| = 8$. Each $\hat{e}_j$ may have an associated duration $d_j$ and/or length $l_j$. Fig. \ref{fig:clickEvents} shows a schematic to illustrate this; the clickstream logs here would generate: Pl, with $l_1 = (t_2 - t_1) \cdot r$ and $d_1 = t_2 - t_1$; Sf, with $l_2 = p_2 - p'_2$; Pl, with $l_3 = p_3 - p_2$ and $d_3 = t_3 - t_2$; Pa, with $d_4 = t_4 - t_3$; Sb, with $l_5 = p'_4 - p_4$. Note that we are inserting Pl and Pa events in-between other events, to incorporate the state of the video player during those times. This critical information is not captured through only the events in the raw data, and has been neglected in other work (\eg in \cite{sinha2014your}).

\parab{Denoising clickstreams}. It is important to remove noise in the video-watching trajectories associated with unintentional user behavior. We handle two cases of events separately:

\para{Combining events}: We combine repeated, sequential events that occur within a short duration (5 sec) of one another, since this pattern indicates that the user was adjusting to a final state. This is a common occurrence with forward (Sf) and backward (Sb) skips, where a user repeats the same action numerous times in a few seconds in seeking the final position; this should be treated as a single skip to the final location. Similarly, a series of Rf or Rs events may occur in close proximity, indicating that the user was in the process of adjusting the rate to the final value. Formally, if there is a sequence of clicks $E_i, E_{i+1}, ..., E_{i+K}$ for which $e_i = e_{i+1} = \cdots = e_{i+K}$ and $t_{i+k+1} - t_{i+k} < 5 \; \forall k \in \{0,...,K-1\}$, then we use $E'_i = \langle e_i, p_{i+K}, t_{i}, s_{i+K}, r_{i+K} \rangle$ in place of $E_i, E_{i+1}, ..., E_{i+K}$.

\para{Discounting intervals}: We identify two instances in which play (Pl) and pause (Pa) events should not be inserted between $E_i$ and $E_{i+1}$. First is if $E_i$ and $E_{i+1}$ occur on two different videos; here, there is no continuity as the user must have exited the first video and then opened the second one. Second is if the duration $t_{i+1} - t_i$ is extremely long; in this case, it is likely that the user was engaging in some off-task behavior during this time. If $s_i = {\tt paused}$, the threshold on the duration is set to 20 min (as in \cite{wang2013you} for web inactivity); if $s_i = {\tt play}$, then the threshold is set to the length of the video.

\subsubsection{Event lengths}
\label{sssec:evtLen}
\begin{figure}
\vspace{-0.05in}
\centering
\subfloat[Boxplots of the distributions for each dataset.]{
\includegraphics[scale=0.32]{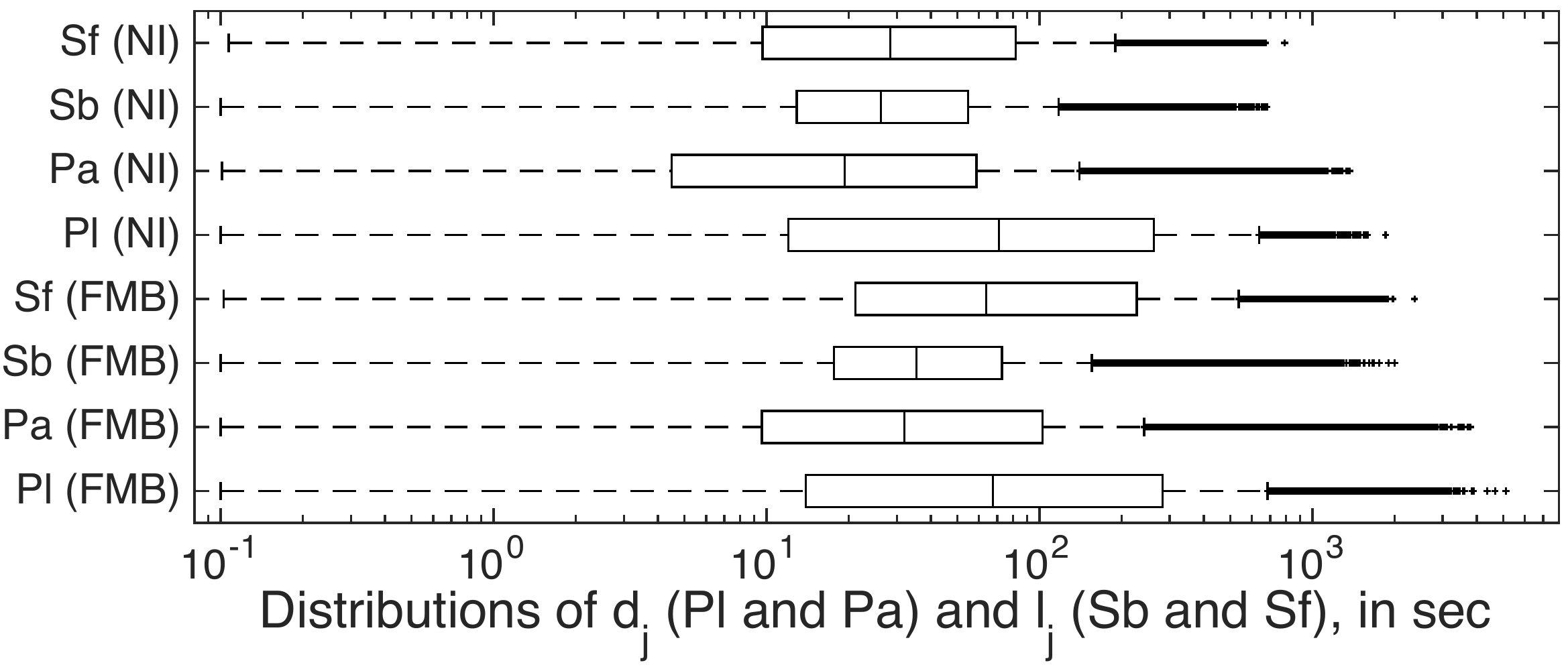}}
\bgroup
\def\arraystretch{1.05}
\subfloat[Tabulated statistics for the distributions.]{
{\small
\begin{tabular}{c|cccccc}
Dataset & Event & Size & Frac & $Q_1$ & $Q_2$ & $Q_3$ \\ \hline
\multirow{4}{*}{`FMB'} & Pl & 112.7K & 53\% & 13.9 & 67.5 & 282.4 \\
 & Pa & 51.2K & 24\% & 9.6 & 31.9 & 102.4 \\
  & Sb & 29.4K & 14\% & 17.7 & 35.4 & 72.7 \\
  & Sf & 18.2K & 8.6\% & 21.2 & 63.7 & 227.2 \\ \hline
\multirow{4}{*}{`NI'} & Pl & 103.5K & 58\% & 12.0 & 71.0 & 262.6 \\
 & Pa & 46.4K & 26\% & 4.5 & 19.3 & 58.8 \\
  & Sb & 17.8K & 10\% & 12.9 & 26.2 & 54.7 \\
  & Sf & 10.7K & 6.0\% & 9.6 & 28.4 & 81.7 \\
\end{tabular}}}
\egroup
\caption{Distribution of the lengths for four events across both `NI' and `FMB'. For Pl and Pa, this represents the time elapsed before the next event, and for Sb and Sf, this is the distance of the skip.}
\label{fig:evtStats}
\vspace{-0.2in}
\end{figure}

We now look to discretize the length $l_j$ and duration $d_j$ of the events. Fig. \ref{fig:evtStats}(a) gives the boxplots of the event distributions from each course. $d_j$ for Pl and Pa is shown, and we depict $l_j$ for Sb and Sf (we show only values that are at least 0.1 sec). Basic statistics of each distribution are also given in Fig. \ref{fig:evtStats}(b); specifically, the three quartiles $Q_1$, $Q_2$, and $Q_3$ are shown,\footnote{By definition, quartiles separate data in increments of 25\%.} as are the number of events for each distribution (Size) and the respective fractions (Frac).

We make three observations in comparing the distributions. In each case, we employed a Wilcoxon Rank Sum (WRS) \cite{sheskin2003handbook} test for the null hypothesis that there was no difference between the distributions for each dataset overall, and report the p-values ($p$) from those tests:\footnote{We use the WRS test because Shapiro-Wilk tests detected significant departures from normality for each of the distributions.} 

\para{(i) `FMB' has longer events}: The distributions for each event are shifted to the right for `FMB' relative to those for `NI', meaning that `FMB' tends to have longer events. In each of the four cases (Pl, Pa, Sb, and Sf), the p-values ($p$) were highly significant ($p \approx 0$). The fact that Pa is longer for `FMB' is consistent with that content being more difficult.

\para{(ii) Sf is longer than Sb}: The distribution of Sf is shifted to the right relative to Sb for both `FMB' and `NI' ($p < $1E-6). This indicates that when students skip forward, they tend to pass more material than they revise when skipping back. Sb also occurs more frequently than Sf for both courses.

\para{(iii) Pl is longer than Pa}: The distributions for play and pause in both datasets indicate that users tend to stay in the {\tt playing} state longer than {\tt paused} ($p \approx 0$). This is stronger in the case of `NI', which is again consistent with the fact that the `FMB' material is more difficult.

\parab{Event intervals}. Clearly, $l_j$ and $d_j$ can vary substantially between events and datasets. To account for this relative variation, we will use the four intervals in-between the three quartiles for each event (given in Fig. \ref{fig:evtStats}(b)) to discretize the lengths. We specify three cases:

\noindent (i) $\hat{e}_j \in \{\mbox{Sb}, \mbox{Sf}\}$: When the event is a skip, we map it to $\langle \hat{e}_j \; q_j \rangle$, where $q_j \in \{1, 2, 3, 4\}$ is chosen such that $l_j \in [Q_{q_{j}-1}, Q_{q_j})$, with $Q_0 = 0$ and $Q_4 = \infty$. For example, suppose that event $E_i$ is such that $\hat{e}_j = \mbox{Sb}$ and $l_j = 20$ sec. In either course, this would be mapped to {\tt Sb2}.

\noindent (ii) $\hat{e}_j = \mbox{Pa}$: In this case, the mapping works the same as the previous, except $q_j$ is chosen based on $d_j$ instead.

\noindent (iii) $\hat{e}_j = \mbox{Pl}$: Two long duration play events could still have different qualitative interpretations.\footnote{The other events do not have this issue since they are not related to processing new, incoming information.} To account for this, when $\hat{e}_j = \mbox{Pl}$, we map it to $\langle \hat{e}_j \; q_{j,1} \;\; \hat{e}_j \; q_{j,2} \; \cdots \; \hat{e}_j \; q_{j,K} \rangle$, where $q_{j,k} \in \{1, 2, 3\}$ for $k = 1,...,K$ is chosen according to:
\begin{equation*}
q_{j,k} = \begin{cases} 3, & d_j - \delta_{j,k} > Q_3 \\ {\arg \min}_{q_{j,K}} (d_j - \delta_{j,K} \leq Q_{q_{j,K}}), & \mbox{otherwise}, \end{cases}
\end{equation*}
with $\delta_{j,k} = \sum_{k' = 1}^{k - 1} Q_{q_{j,k'}}$ at each step. For example, suppose an event is Pl with $d_j = 550$ sec. For the quartiles in `NI', this would be mapped to {\tt Pl3 Pl3 Pl2}.

\subsubsection{Event-type sequence specification}
\label{sssec:fullEvtSpec}
Let $\mathcal{S} = \{{\tt Pl1}, {\tt Pl2},$ ${\tt Pl3}, {\tt Pa1}, ..., {\tt Pa4}, {\tt Sb1}, ..., {\tt Sb4}, {\tt Sf1}, ..., {\tt Sf4}, {\tt Rf}, {\tt Rs},$ ${\tt Rd} \}$ be the set of $|\mathcal{S}| = 18$ events (with quantized lengths). For each UV Pair, we encode the clickstream log $E_1, ..., E_n$ as $S = (s_1, s_2, ...,$ $s_{n'})$ where each $s_j \in \mathcal{S}$ is chosen according to the specifications in Sec. \ref{sssec:evtLen}. As we will see in Sec. \ref{sec:motifs}, using this alphabet that incorporates event types and lengths allows us to obtain insights that cannot be gleaned with events alone.

For comparison, we will refer to an event with length 1 as ``short,'' 2 as ``medium,'' 3 as ``medium-long,'' and 4 as ``long.''

\section{Motifs of Video-Watching}
\label{sec:motifs}
Using the event-type specification, we identify short, recurring sub-sequences within user behavior, \ie behavioral \textit{motifs}. As we will see in Sec. \ref{ssec:motifRes}, these motifs capture fundamental video watching characteristics of students such as reflecting on or revising material. We will also see that some of these motifs are significantly associated with performance.

\subsection{Motif Extraction}
\label{ssec:motifExt}
We make use of the MEME Suite software package \cite{bailey2009meme} for motif extraction. MEME has been applied in bioinformatics for motif identification in sequences of nucleotides and amino acids. We turn meme MEME to be applicable in our setting.

\parab{Model and algorithm}. The underlying algorithm is based on a probabilistic mixture model, where the key assumption is that each subsequence is generated by one of two components: a position-\textit{dependent} motif model, or a position-\textit{independent} background model. Under the motif model, each position $j$ in a motif is described by a multinomial distribution, which specifies the probability of each character (\ie each $s \in \mathcal{S}$ from Sec. \ref{sssec:fullEvtSpec}) occurring at $j$. The background model is a multinomial distribution specifying the probability of each character occurring, independent of the positions; we employ the standard background of a $0$-order Markov Chain. A latent variable is assumed that specifies the probability of a motif occurrence starting at each position in a given sequence \cite{bailey2009meme}.

Motif extraction is formulated as a maximum likelihood estimation over this model, and an expectation-maximization (EM) based algorithm is used to maximize the expectation of the (joint) likelihood of the mixture model given both the data (\ie the sequences) and the latent variables. We use the standard dirichlet prior based on character frequencies for EM.

\parab{Extraction}. Each UV-Pair's clickstream sequence is encoded using the $24$-character protein alphabet  \cite{bailey2009meme}. To do this, we choose the first $18$ non-ambiguous characters $\mathcal{F}$, and then specify a 1:1 mapping $\mathcal{S} \leftrightarrow \mathcal{F}$. Whereas other work has focused on a single motif width (\eg at $4$ in \cite{sinha2014your}), we extract those of widths $w \in \{4,...,10\}$ from our datasets, with E-values (see below) at most $0.05$; we will see that both long and short motifs can be insightful (see Fig. \ref{fig:motifs}).

For each motif, we obtain its E-value, and its position specific probability matrix (PSPM):

\para{E-value}: The $E$-value judges overall significance. It is defined as the fraction of motifs (with the same width and occurrences) that would have higher log likelihood ratio if the sequences had been generated according to the background model.

\para{PSPM}: This gives the fraction of times that each character appears in each position of the motif, taken over all sightings of the motif in the dataset. In the following, denote the PSPM for a motif by $\mathbf{P} = [p_{i,j}]$, where $p_{i,j}$ is the fraction of times event $j$ occurs at position $i$.

\parab{Representation}. At each position $i$, we consider all events $j$ with $p_{i,j} \geq 0.25$.\footnote{With 18 different events, a threshold of 25\% is roughly 5 times the expected occurrence from a purely random selection of events.} Formally, let $\mathcal{A}_i$ be the sequence of indices into the event set $\mathcal{S}$ for $i$, arranged such that $p_{i,\mathcal{A}_i(k)} \geq p_{i,\mathcal{A}_i(k+1)}$ and $p_{i,\mathcal{A}_i(k+1)} \geq 0.25 \; \forall k$. Then, there are three cases on the way $i$ is represented: if $|\mathcal{A}_i| >1$, $i$ is represented  as $[\mathcal{S}_{A_i(1)} \; \mathcal{S}_{A_i(2)} \; \cdots]$; if $|\mathcal{A}_i| = 1$, then the square brackets are omitted, with just $\mathcal{S}_{A_i}$ displayed; if $A_i = \emptyset$, then $i$ is displayed as `$\star$' to indicate that this position was taken by a variety of events, none of which occurred even 25\% of the time. For example, the sequence {\tt [Pl2 Pl3] Pa1 $\star$ [Sf1 Sf2 Sf4]} is of length 4, with the first position being either {\tt Pl2} or {\tt Pl3} at least 50\% of the time ({\tt Pl2} at least as often as {\tt Pl3}), the second position being {\tt Pa1} at least 25\% of the time, the third position being any event, and the last being either {\tt Sf}, {\tt Sf2}, or {\tt Sf4} at least 75\% of the time.

\parab{Motif support}. For each motif, we obtain the fraction of sequences (FS) in which it occurs, \ie its support across sequences, as well as the number of videos it appears in. We also obtain FS0 and FS1 as the fraction of non-CFA and CFA sequences in which the motif appears, respectively.

\para{Significance test}: We determine whether there is a significant difference in the support of a motif across the CFA and non-CFA classes by running a two-sample test for proportions \cite{sheskin2003handbook} for the null hypothesis $H_0$ that $\mbox{FS1} = \mbox{FS0}$, with an alternative hypothesis $H_1$ that $\mbox{FS1} \neq \mbox{FS0}$. If the two-sided $p$-value ($p$) for this test is low enough ($\leq 0.05$), then the difference between the supports is significant, \ie the motif is found in the class with higher support significantly more often. We also compute the estimated probability of success $\hat{p} \in [0,1]$ (\ie of a CFA submission) for a sequence containing the motif, from the midpoint of the confidence interval returned by this test.

\begin{figure}
\vspace{-0.2in}
\centering
\subfloat[`FMB']{
\includegraphics[scale=0.24]{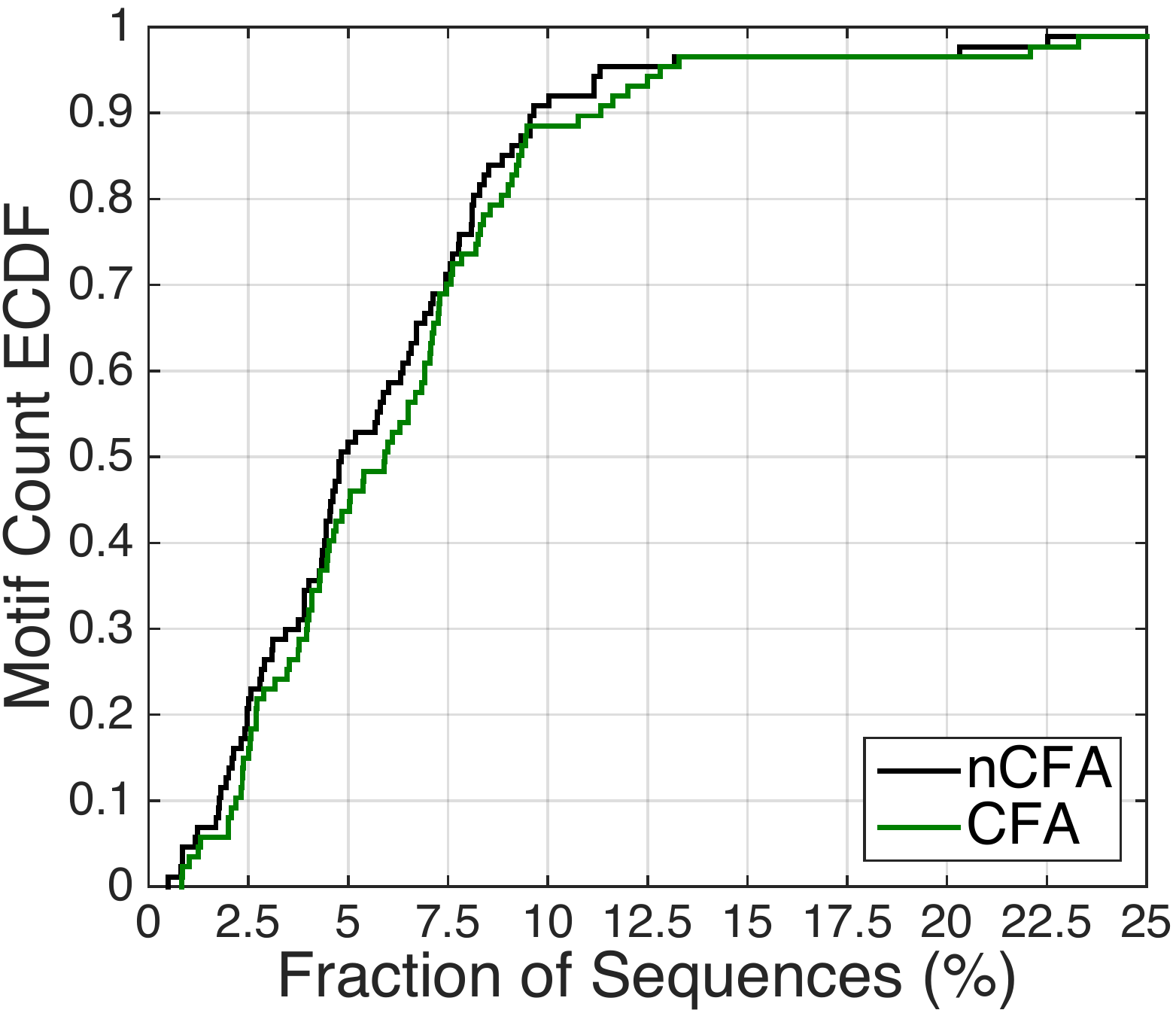}} \hspace{0.05in}
\subfloat[`NI']{
\includegraphics[scale=0.24]{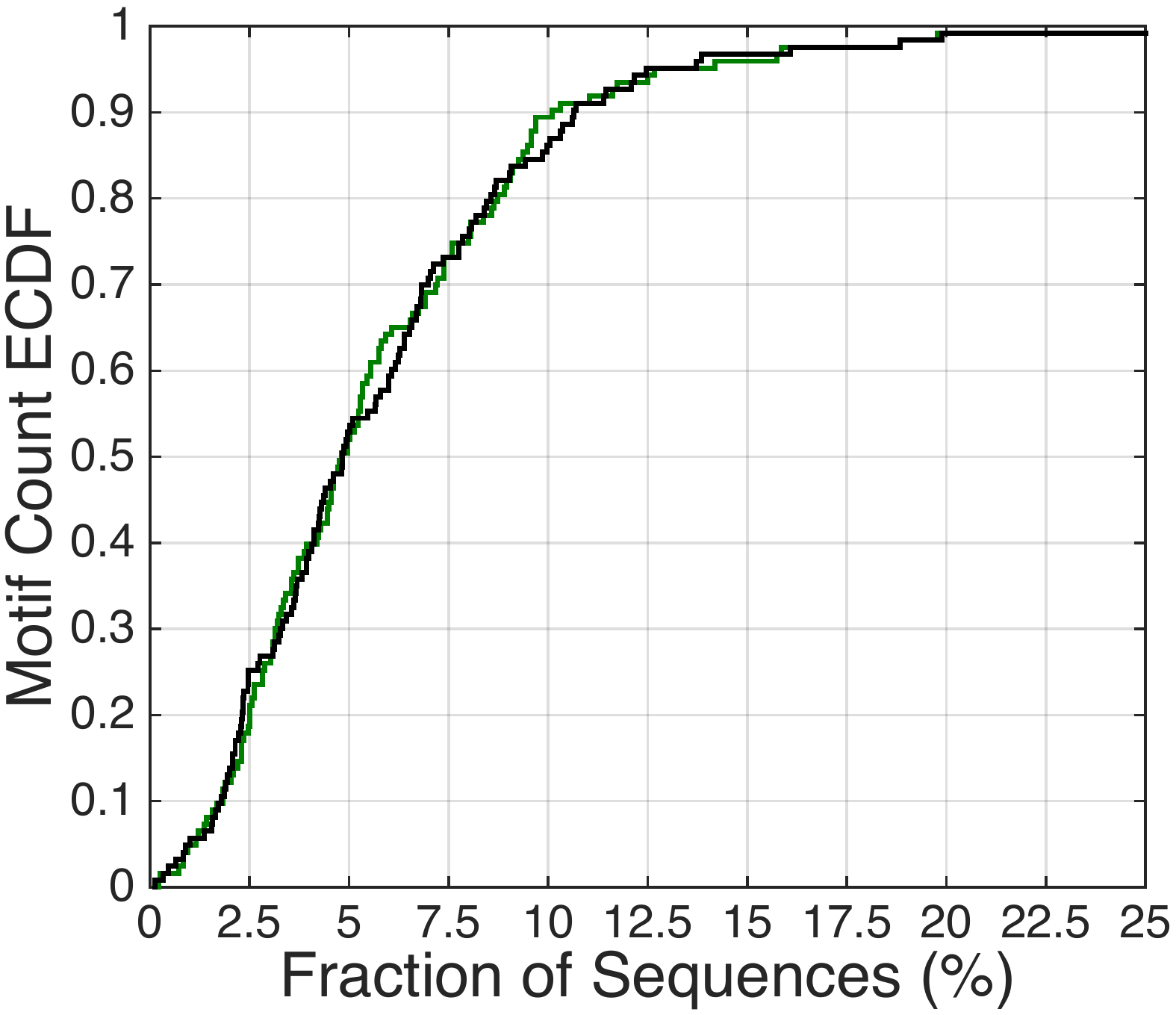}}
\caption{ECDFs of the number of sequences that each motif appears in, across both CFA and non-CFA. The supports are consistent across both groups.}
\label{fig:mot_s_cdf}
\vspace{-0.25in}
\end{figure}
\begin{figure}
\centering
\subfloat[`FMB']{
\includegraphics[scale=0.24]{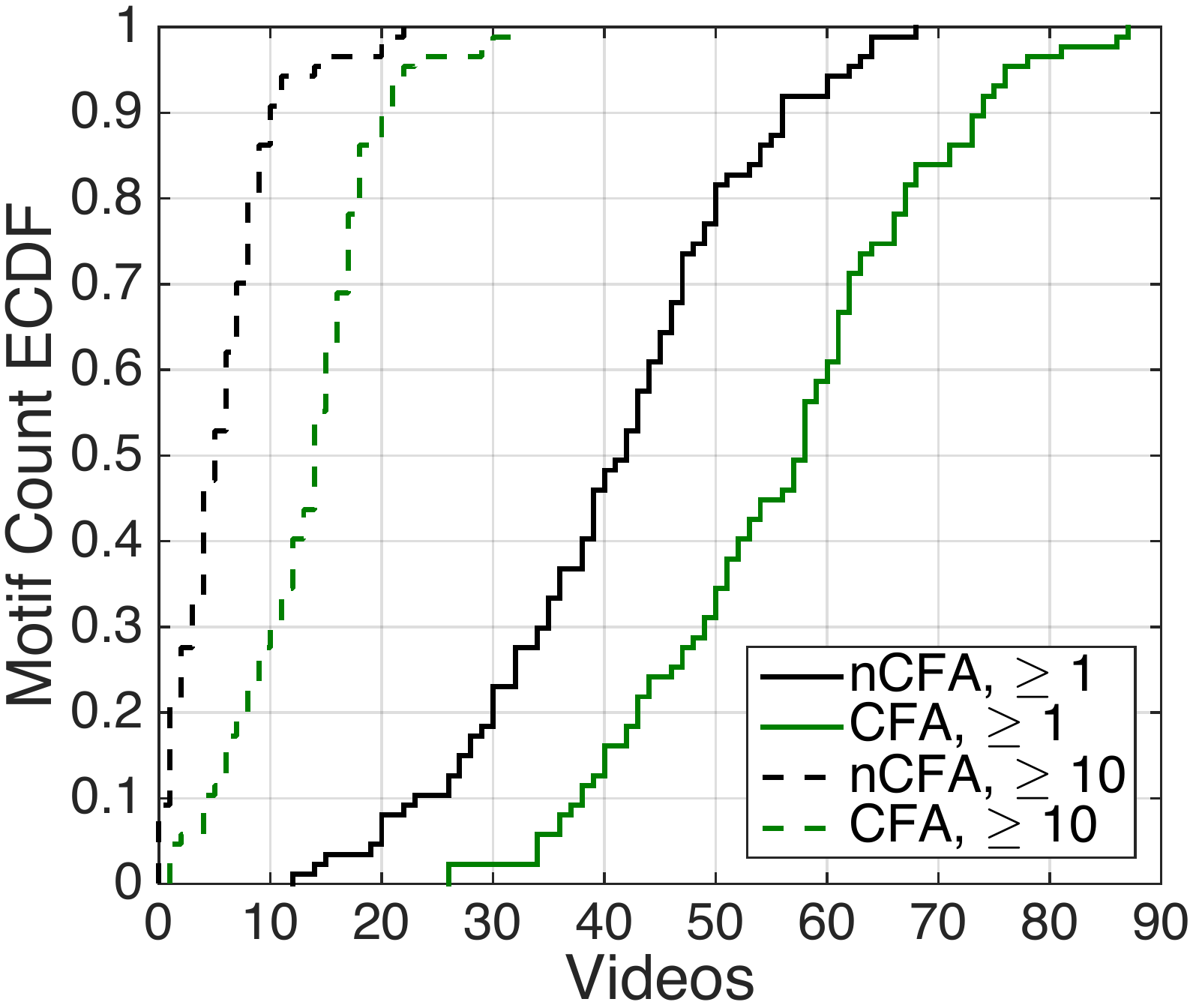}} \hspace{0.05in}
\subfloat[`NI']{
\includegraphics[scale=0.24]{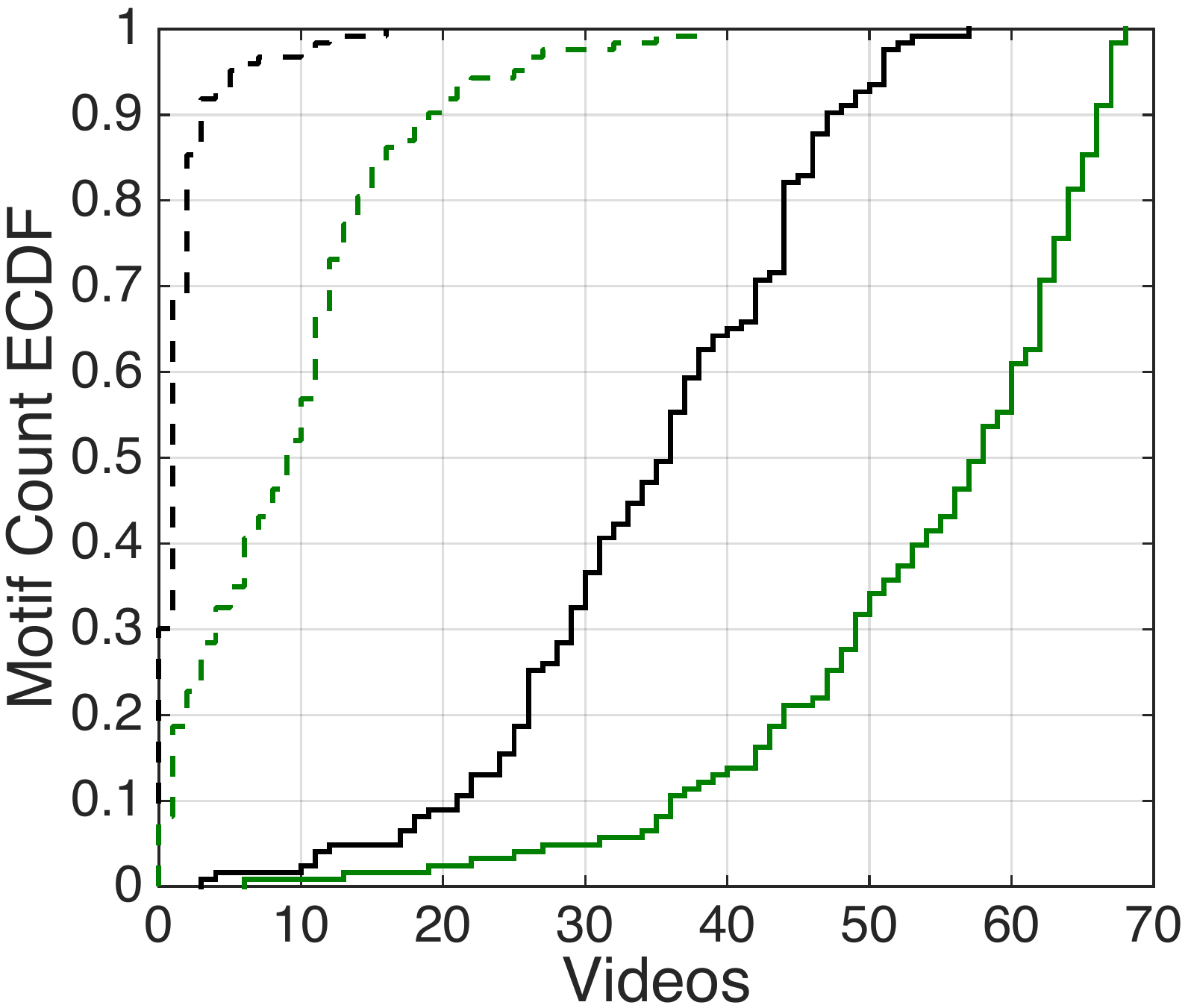}}
\caption{ECDFs of the number of videos that each motif appears in, across both CFA and non-CFA sequences. CFA sequences have a higher support for motifs across videos.}
\label{fig:mot_v_cdf}
\vspace{-0.2in}
\end{figure}

\begin{figure*}[t]
\vspace{-0.1in}
\centering
\bgroup
\def\arraystretch{1.05}
\subfloat[Motifs for `FMB'.]{
\footnotesize
\begin{tabular}{c|cc|cc|cccc}
Group &  & Motif & E-value & FS(\%) & FS0(\%) & FS1(\%) & $\hat{p}$(\%) & p-value \\ \hline
\multirow{4}{*}{Pa} & I &\mbox{[Pl2 Pl3] [Pa4 Pa3] [Pl2 Pl1] [Pa2 Pa3] Pl2 Pa3 Pl2 [Pa2 Pa3] Pl3} 	& 5.3E-64 & 28.5 & 26.2 & 29.5 & 53.3 & 4.6E-4$^{**}$ \\
 & II & \mbox{Pl2 Pa4 Pl2 Pa4} & 1.5E-06 & 13.2 & 13.2 & 13.3 & 50.1 & 0.901 \\
  & III & \mbox{[Pa1 Pa3] Pl1 [Pa2 Pa1] Pl1 [Pa1 Pa2] [Pl1 Pl2] [Pa1 Pa2] [Pl1 Pl2]} & $\approx 0$ & 12.1 & 11.3 & 12.5 & 51.2 & 0.0745 \\
 & IV &\mbox{[Pl1 Pl3 Pl2] Pa4 Pl2 Pa3 [Pl2 Pl1] $\star$ Pl2 $\star$ [Pl3 Pl1] Pa3} & 1.5E-15 & 10.9 & 9.3 & 11.6 & 52.3 & 3.9E-4$^{**}$ \\ \hline
\multirow{3}{*}{Sb} &I &  \mbox{Sb3 [Pl2 Pl1] [Sb2 Sb3] Pl2 Sb2 Pl2 [Sb2 Sb3] [Pl3 Pl2]} & 6.0E-245 & 10.2 & 8.84 & 10.8 & 51.9 & 2.0E-3$^{**}$ \\
 & II &  \mbox{Pl3 Sb3 Pl2 [Sb3 Sb2] Pl3} & 6.7E-40 & 8.92 & 8.11 & 9.28 & 51.2 & 0.048$^{*}$ \\
 & III & \mbox{Pl2 Sb2 [Pl1 Pl2] [Sb2 Sb3] Pl3} & 0.044 & 7.79 & 6.57 & 8.32 & 51.7 & 1.6E-3$^{**}$ \\ \hline
\multirow{2}{*}{Sf} & I & \mbox{Pl2 Sf3 [Pl1 Pl2] Sf2 [Pl1 Pl2] Sf1 [Pl2 Pl1] [Sf2 Sf1]} & $\approx 0$ & 9.46 & 10.03 & 9.22 & 49.2 & 0.186 \\
 & II & \mbox{[Pl2 Pl1] [Sf2 Sf3] Pl1 [Sf3 Sf2]} & $\approx 0$ & 6.42 & 7.41 & 5.98 & 48.6 & 4.95E-3$^{**}$ \\ \hline
\multirow{2}{*}{Rf} & I & \mbox{Pl3 [Rf Rd] [Pl2 Pl1] Rf [Pl3 Pl2] Rf} & $\approx 0$ & 4.55 & 3.89 & 4.84 & 50.9 & 0.0295$^{*}$ \\
 & II & \mbox{Rf Rd [Pl1 Pl2] Rf Pl3} & 1.2E-70 & 1.77 & 1.22 & 2.00 & 50.8 & 4.7E-3$^{**}$ \\ \hline
\end{tabular}} \\
\vspace{-0.05in}
\subfloat[Motifs for `NI'.]{
\footnotesize
\begin{tabular}{c|cc|cc|cccc}
Group & & Motif & E-value & FS(\%) & FS0(\%) & FS1(\%) & $\hat{p}$(\%) & p-value \\ \hline
\multirow{4}{*}{Pa} &I & \mbox{[Pl2 Pl3] Pa4 [Pl2 Pl3] Pa4 Pl3} & 2E-81 & 26.8 & 27.6 & 26.5 & 48.8 & 0.338 \\
& II &\mbox{Pl2 Pa4 Pl2 Pa4} & 1.8E-44 & 14.3 & 15.9 & 13.7 & 47.9 & 0.0233$^{*}$ \\
&III & \mbox{Pl2 Pa4 Pl2 Pa3 Pl3} & 3.2E-19 & 11.8 & 11.0 & 12.1 & 51.0 & 0.241 \\
&IV & \mbox{Pl1 Pa1 Pl1 Pa1 Pl1 Pa1 [Pl1 Pl3]} & $\approx 0$ & 11.7 & 12.7 & 11.4 & 48.7 & 0.145 \\ \hline
\multirow{2}{*}{Sb} &I & \mbox{[Sb3 Sb4] [Pl2 Pl3] [Sb3 Sb2] Pl2 [Sb3 Sb2] [Pl3 Pl2]} & 9.1E-191 & 9.2 & 8.6 & 9.4 & 50.8 & 0.291 \\
&II & \mbox{Sb3 Pl2 Sb2 [Pl2 Pl1] Sb2 Pl2 [Sb3 Sb2] [Pl3 Pl2]} & 2.2E-125 & 5.3 & 4.2 & 5.7 & 51.5 & 0.014$^{*}$ \\ \hline
\multirow{3}{*}{Sf} &I & \mbox{[Pl3 Pl1] [Sf3 Sf4] [Pl1 Pl2] [Sf4 Sf3 Sf2] [Pl1 Pl2] [Sf3 Sf4]} & 6.6E-100 & 7.8 & 8.9 & 7.4 & 48.4 & 0.0279$^{*}$ \\
&II & \mbox{Pl2 [Sf3 Sf2] Pl1 [Sf3 Sf2]} & 1.1E-248 & 7.7 & 9.7 & 7.0 & 47.4 & 2.2E-4$^{**}$ \\
& III & \mbox{Pl2 Sf3 [Pl1 Pl2] [Sf3 Sf2] [Pl3 Sb1]} & 2.7E-3 & 6.3 & 7.2 & 6.0 & 48.8 & 0.0598 \\ \hline
\multirow{2}{*}{Rf} & I & \mbox{Rf [Pl2 Pl1] Rf [Pl3 Pl2]} & $\approx 0$ & 2.3 & 3.5 & 1.9 & 48.4 & 9E-5$^{**}$ \\
 & II & \mbox{[Rf Rs] [Pl2 Pl1] Rd [Pl2 Pl3] Rf} & 7.3E-16 & 2.5 & 3.1 & 2.3 & 49.2 & 0.064 \\ \hline
\end{tabular}}
\egroup
\vspace{-0.05in}
\caption{Representative sample of motifs identified for each course. Each motif is grouped by the dominant event it contains outside of Pl. FS is the fraction of sequences over both CFA and non-CFA, while FS0 and FS1 are for the separate cases. $\hat{p}$ is the estimated probability of success (CFA) if a sequence contains the motif, and the p-value ($p$) is the significance of $\hat{p}$ ($^{*}$ indicates $p \leq 0.05$, and $^{**}$ is for $p \leq 0.01$).}
\vspace{-0.225in}
\label{fig:motifs}
\end{figure*}

\subsection{Results}
\label{ssec:motifRes}
We obtained 87 and 123 motifs from `FMB' and `NI', respectively, which are the subject of the following analysis.

\parab{Motif overview}. We first analyze how the motif supports vary across sequences and videos. Overall, we find that the motifs are reasonably supported across sequences and videos on average, for both CFA and non-CFA in each course.

\para{Sequences}: In Fig. \ref{fig:mot_s_cdf}, we plot the Empirical CDF (ECDF) of the fraction of sequences that each motif appears in, for both CFA and non-CFA. The supports are similar across these groups: for `FMB', each motif appears in 5.9\% of the non-CFA sequences on average, and 6.5\% of the CFA; for `NI', this is 5.8\% for CFA, and 4.2\% of the non-CFA. In both courses, the motifs with largest support (first row in Fig. \ref{fig:motifs}(a) and (b)) appear in $> 25\%$ of the sequences.

\para{Videos}: Fig. \ref{fig:mot_v_cdf} gives the ECDF of the number of videos that each motif occurred in at least once and at least 10 times. Overall, CFA has higher support than non-CFA over videos. We also see that the supports decrease for higher thresholds, \eg for `FMB' in (a), while the top 20\% of the motifs appear in at least 67 videos for CFA, this drops to only 18 videos considering at least 10 occurrences.

\subsubsection{Individual motifs}
\label{sssec:indivMotif}
We inspect patterns in the most significant of the 210 extracted motifs. This list is obtained by applying the following procedure. First, noticing that all motifs contain Pl events, we group them into categories based on the most recurring alternate event, leading to 4 groups. Then, within each category, we consider each motif that either (i) has one of the top-10 highest supports or (ii) has a significant $p$ ($\leq 0.05$) comparing CFA and non-CFA supports. Finally, if one motif is a subsequence of another, then we remove the one that has lower support or is less significant.

This yields 19 and 21 motifs for `FMB' and `NI', respectively. In Fig. \ref{fig:motifs}, we give the representative sample of these 40 that are mentioned in the following discussion. Note that we have grouped each motif by the most frequent type of event that it contains aside from {\tt play}. Also, each motif is assigned an ID consisting of its group and number (\eg Pa II in `FMB' is motif {\tt Pl2 Pa4 Pl2 Pa4}).

\parab{Overview}. The motifs exhibit many similar structural attributes, which occur in spite of the fact that the encoding quantiles are different for each event and course (see Fig. \ref{fig:evtStats}). Also, since MEME finds ungapped motifs (\ie those existing as exact matches in the data, without a separate layer of similarity matching), these identified behaviors exist exactly in the sequences, contrary to other work \cite{sinha2014your} which has resorted to approximate string searching. Interestingly, we find that the motif with highest support in each group also tend to have the longest length (average of $7.5$ over all groups with at least two motifs). Also, we find that the motifs in the Pa group have the largest supports (FS) overall ($\geq 10\%$  mostly), which is consistent with the fact that there are less {\tt skip} and {\tt ratechange} events in the datasets (see Fig. \ref{fig:evtStats}(b)).

We present our most interesting observations for each group:

\parab{Reflecting (Pa)}. The occurrence of play together with pause indicates that lectures are generally thought-provoking, causing students to \textit{reflect on the material they just saw}. In both courses, the events forming the motifs in this group cover the entire range from short to medium-long plays ({\tt Pl1} -- {\tt Pl3}) interspersed with short to long pauses ({\tt Pa1} -- {\tt Pa4}).

The motif with the highest support in `FMB' -- Pa I -- can be viewed as a long sequence of medium to medium-long plays with medium-long to long pauses in-between, also characteristic of Pa IV in `FMB'. This behavior occurs more often in the CFA group in both cases ($p < 0.01$). Motifs Pa III in `FMB' and Pa IV in `NI' are long sequences too, but consist of short to medium plays followed by short to medium pauses and do not distinguish between CFA and non-CFA ($p > 0.07$). Motifs Pa II in `FMB' and Pa I in `NI' are shorter sequences, with medium to medium-long plays followed by long pauses, and also do not differentiate between the groups ($p > 0.33$).

In `NI', the Pa group exhibits less significance in $p$-values. For this reason, we do not draw conclusions on differences between CFA and non-CFA from these sequences.

\parab{Revising (Sb)}. From the six motifs in the Sb group, we identify two interesting, recurring subsequences: {\tt Pl2 Sb3 Pl2 Sb3} (Sb I and II for `FMB', and Sb I for `NI'), and {\tt Pl2 Sb2 Pl2 Sb2} (Sb III for `FMB' and Sb II for `NI'). Roughly speaking, each of these is associated with \textit{playing for a length of video, and then revising some or all of that content}. To see this, consider the ranges of Pl and Sb from Fig. \ref{fig:evtStats} associated with these subsequences: {\tt Pl2} covers 14 to 68 sec for `FMB', and {\tt Sb2} to {\tt Sb3} covers 18 to 73 sec; for `NI', these ranges are 12 to 71 sec and 13 to 55 sec. The play and skip ranges are closely overlapping in each case. Taking the extreme ends of each range, they are associated with skipping back anywhere from 1 min below the starting play point to 50 sec after it,\footnote{We assume a default playback rate as an approximation.} which are local considering the video lengths. This characteristic of local revising is further seen in that {\tt Sb4}, a \textit{long} skip back, does not appear in these motifs. Note that \textit{4 of the 5 motifs containing these behaviors are significantly associated with CFA} ($p < 0.05$).

We also considered the number of skip backs originating at each video position across all UV Pairs. We find that the largest origination point of these events is at the end of videos. In particular, out of all Sb events, those originating within 10 sec of the videos' end are 16\% and 13\% of the total for non-CFA and CFA in `FMB'. As a reference, if we take the highest location of Sb for each video outside of the last 10 sec, these constitute 4.5\% and 3.6\% of the total for non-CFA and CFA. This, combined with the motifs suggesting revision when Sb occurs, implies that those students who are revising \textit{multiple times} before answering a quiz have a higher chance of success.

\parab{Skimming (Sf)}. In both of the courses, the motifs in the Sf group are primarily medium to long skips forward with short to medium plays in-between. Further, the skips are longer than the plays occurring before and after; comparing the lengths of Pl and Sf events in Fig. \ref{fig:evtStats}, we see that for both courses, range $Q_j$ to $Q_{j+1}$ for Sf is always larger than $Q_{j-1}$ to $Q_j$ for Pl. This recurring behavior in the Sf group can then be interpreted as \textit{skimming through the material quickly} with less exposure to the material. We find that 3 of these 5 motifs are significant in favor of non-CFA ($p \leq 0.03$). We contrast this to a finding in \cite{brinton2015prediction}, where the total number of skip forwards in a sequence was not found to be associated with either group. This underscores the utility of considering the clickstream sequences, rather than computing aggregate quantities to summarize them.

While Sb and Sf occurring together in a motif (\eg Sf III in `NI') can possibly be interpreted as \textit{skipping forward with caution}, we find that this is also close to being significant in favor of non-CFA ($p < 0.06$).

\parab{Speeding (Rf)}. Referring to Rf in `FMB', motifs I and II indicate that viewing the material at a faster than default rate, \ie \textit{speeding}, is more often associated with the CFA class than not ($p < 0.03$). With these motifs, learners also return to the default rate (Rd), indicating they are \textit{slowing down for important content}. To this point, in `FMB', we see no significant motifs for slower than default rates; however, one does exist in `NI' (Rf II). Also, Rf II in `NI' is significantly associated with non-CFA ($p = $ 9E-5), which could indicate that a faster rate is harmful in this course.

\subsubsection{Key messages}
Overall, we draw a few conclusions.

\parab{Motif groups}. There are four main groups:

\para{Reflecting}: Pausing to reflect on material repeatedly is the most commonly recurring behavior. If the time spent reflecting is not \textit{too} long, but longer than the time spent watching, then a positive outcome is most likely (in `FMB').

\para{Revising}: Repeated revision of the material suggests students will gain a better understanding of the content.

\para{Skimming}: Skimming through material quickly, even with caution, is costly in terms of knowledge gained.

\para{Speeding}: Students who watch the videos at a faster than default rate may already be familiar with the material, leading to a correct answer (in `FMB'). They also may slow to the default if they sense unfamiliar material.

\parab{Significance of associations}. For each motif, the identified association with CFA or non-CFA is particularly important, because in many cases either would be intuitive. For example, a revising motif could presumably come from a student reinforcing material prior to the quiz (in line with CFA) or from excess confusion caused by the material (in line with non-CFA), but the results indicate the former is more likely. As another example, skimming could come from a student perceiving familiarity with the content in a video, which could intuitively be either a correct (in line with CFA) or an incorrect (in line with non-CFA) perception, but results favor the latter.

\parab{Importance of lengths}. We emphasize the importance of having included the lengths, in addition to the events, in our framework from Sec. \ref{sssec:fullEvtSpec} in order to make these conclusions. For instance, the sequence {\tt Pl Sb Pl Sb} identified in \cite{sinha2014your} cannot be associated with revising, because it is not clear how far back the student has skipped relative to having played in-between. In the same way, {\tt Pl Sf Pl Sf} cannot be concluded as skimming, because the lengths of {\tt play} and {\tt skip} are not indicated in the model. Also, even small changes in the motif lengths can affect significance (\eg in `FMB', while Pa I is associated with CFA, Pa II is not).

Clickstream motifs are useful in studying learning behavior, and that they can be significantly related to performance. In terms of using them to model behavior for CFA prediction, however, there are two drawbacks. First, while the supports are reasonable considering these are rather long subsequences, none of the motifs appear in a majority of the sequences (max 28.5\%). Second, none of the $\hat{p}$ success estimates deviate substantially from 50\% (max 3.3\%). Hence, we will now turn to an alternate clickstream sequence representation which is more applicable to CFA prediction. Nonetheless, some of the conclusions here will guide our modeling choices.

\section{Model of Position Sequence}
\label{sec:positions}
In this section, we will formalize a position-based sequence representation, which factors in the location in the videos that a student visited. Then, we will preset CFA models based on this framework, which will be evaluated in Sec. \ref{sec:modelEval}.

\subsection{Modeling Framework}
\label{ssec:modelOverview}

\subsubsection{Definitions}
Let $v \in \mathcal{V}$ denote video $v$ in the set of videos $\mathcal{V}$ for a course, indexed chronologically (\ie by release date of the videos).\footnote{Recall from Sec. \ref{ssec:ourCourses} that we define a ``video'' to be all videos for a quiz.} Also, let $c \in \mathcal{C}$ denote class $c$ in the set of binary classes $\mathcal{C} = \{0,1\}$, where $c = 0$ indicates a non-CFA submission and $c = 1$ is CFA. With $u \in \mathcal{U}$ as user $u$ in the set of all users $\mathcal{U}$, we let $\mathcal{U}^v \subset \mathcal{U}$ be the set of users who have a UV Pair for $v$, and $\mathcal{U}^{v,c} \subset \mathcal{U}^v$ be those who fall into class $c$ with respect to their answer submission. For evaluation in Sec. \ref{sec:modelEval}, we will generate training ($\mathcal{U}^{v}_T$) and test ($\mathcal{U}^{v}_E$) sets as subsets of $\mathcal{U}^{v}$; $\mathcal{U}^{v}_T$ and $\mathcal{U}^{v}_E$ are always chosen such that $\mathcal{U}^{v}_T \cap \mathcal{U}^{v}_E = \emptyset$.

\subsubsection{Position-based sequence specification}
We will divide each video into a number of intervals. Let $h_v$ be the length (in sec) of $v$. We define $w_v$ to be the width that partitions $v$ into $N(w_v) = \lfloor h_v / w_v \rfloor$ uniform intervals, such that interval $i \in \mathcal{P}^v(w_v) = \{1,...,N(w_v)\}$ spans the range $[(i - 1) \cdot w_v, \; i \cdot w_v]$. For each UV Pair, we can then model the behavior as a sequence of positions $\mathbf{p}^{u,v} = (\rho_1, \rho_2, ..., \rho_n, ...)$, where $\rho_n \in \mathcal{P}^v(w_v)$ is the index of the $n$th position visited.\footnote{For brevity, we will typically refer to $\mathbf{p}^{u,v}$ as just $\mathbf{p}$, with the understanding that it refers to the UV Pair in question.}

To generate these sequences, we first apply the same denoising procedure described in Sec. \ref{ssec:evSpec} to each event $E_i$. Then, for each UV Pair, starting with $\mathbf{p} = ()$ we do the following:

\begin{enumerate}
\item For $E_1$, add $\lfloor p_1 / w_v \rfloor$ to $\mathbf{p}$.

\item Consider each sequential pair of events $E_i, E_{i+1}, \; i \geq 1$. If the state $s_i = {\tt paused}$, then only $ \lfloor p_{i+1} / w_v \rfloor$ is added to $\mathbf{p}$. But if $s_i = {\tt playing}$, then:
\begin{itemize}
\item If the event $e_i \neq \mbox{Skip}$, then $(\lfloor p_i / w_v \rfloor + 1, ..., \lfloor p_{i+1} / w_v \rfloor - 1, \lfloor p_{i+1} / w_v \rfloor)$ is appended to $\mathbf{p}$.
\item If $e_i = \mbox{Skip}$, then $(\lfloor p_i / w_v \rfloor + 1, ..., \lfloor p'_{i} / w_v \rfloor - 1, \lfloor p'_{i} / w_v \rfloor, \lfloor p_{i+1} / w_v \rfloor$) is appended instead.\footnote{Recall from Sec. \ref{ssec:evSpec} that when $E_i$ is a skip event, $p'_i$ is the position of the video player immediately before the skip.}
\end{itemize}
\end{enumerate}

For example, suppose $h_v = 300$, $w_v = 15$, and a user generates $E_1 = \langle {\tt play}, 0, 0, {\tt playing}, 1.0 \rangle$, $E_{2} = \langle {\tt skip}, 200, 50,$ ${\tt playing}, 1.0 \rangle$, $E_{3} = \langle {\tt ratechange},230, 80,{\tt playing}, 1.25 \rangle$, and $E_{4} = \langle {\tt pause}, 300, 127,$ ${\tt paused}, 1.25 \rangle$ on the video. Then, $\mathbf{p} = (0, 1, 2, 3, 13, 14, 15, 15, 16, ..., 20)$.

\subsubsection{Model factors}
\label{sssec:modelFactors}
There are (at least) three types of information for each $\mathbf{p}^{u,v}$ that could have an effect on performance:

\parab{(1) Positions}. First is the number of times a given position $i \in \mathcal{P}^v(w_v)$ was visited. One would expect these to differ between CFA and non-CFA, because certain parts of videos will be more important to questions. To see this, we can refer to two motif groups which were associated with CFA: \textit{reflecting}, which indicates that these sequences may have more visits to important positions through pausing, and \textit{revising}, which suggests that these sequences may have more visits to positions associated with the questions through repeated revision before answering. Further, the \textit{skimming} motif suggests that non-CFA sequences will have less visits to important positions.

\parab{(2) Transitions}. Second is the number of transitions between the positions, \ie the number of times a given tuple $(i, j)$ is a subsequence of $\mathbf{p}^{u,v}$. Considering each tuple $(\rho_n, \rho_{n+1})$:
\begin{itemize}
\item If $\rho_{n+1} < \rho_n$, then the user had skipped back. We call this a \textit{backward} transition.
\item If $\rho_{n+1} > \rho_n + 1$, then the user had skipped over the material in $(\rho_n, \rho_{n+1})$. This is a \textit{forward} transition.
\item If $\rho_{n+1} = \rho_n + 1$, then the user moved directly to the next position. This is a \textit{direct} transition.
\item If $\rho_{n+1} = \rho_n$, then the user had some event \textit{within} the current position. This is a \textit{repeat} transition.
\end{itemize}
We say that direct and repeat transitions are \textit{local}, whereas backward and forward are \textit{non-local}. As with positions, the transition factors can capture the motif behavior associated with CFA and non-CFA, except in terms of sequences of visits, \eg backward transitions capture the Sb in a revising motif, and forward transitions capture the Sf in a skimming motif.

\parab{(3) Time spent}. The amount of time spent at the different positions. One would expect these times to be indicative of CFA/non-CFA in a similar manner to visit frequencies.

In order to evaluate the benefit of including each of these factors, we will consider three prediction models: Discrete time Positions (DP), which incorporates the number of visits to each position; Discrete time Transitions (DT), which models transitions between positions; and Continuous time Transitions (CT), which factors in inter-arrival times between positions. Each model will be tested on each video separately, allowing us to compare results on a per-video basis in Sec. \ref{sec:modelEval}.

\subsection{Position-Based Modeling}
\label{ssec:posModel}
\parab{Discrete Time Positions (DP)}. For the DP model, video positions are treated as independent events. Let $\mathbf{f}^{v,c} = [f_i]^{v,c} \in [0,1]^{N(w_v)}$ be the probability distribution of visit frequency across positions $i \in \mathcal{P}^v(w_v)$. This is estimated over the UV Pairs in the training set $\mathcal{U}_T^{v,c}$ as
\begin{equation}
f^{v,c}_i = O^{v,c}_i / \sum_j O^{v,c}_j,
\label{eqn:cohortDP}
\end{equation}
where $O^{v,c}_i$ is the number of occurrences of $p_i$ over sequences in $\mathcal{U}_T^{v,c}$, \ie  $O^{v,c}_i = \sum_{u \in \mathcal{U}_T^{v,c}} \sum_{n} \mathbb{I}_{\{\rho_n = i\}}$.

We test the ability of this model to identify which class each $u \in \mathcal{U}^v_E$ belongs to. For this purpose, we compute the likelihood of observing $\mathbf{p}$ on video $v$ to be in $c$, given $\mathbf{f}^{v,c}$, as
\begin{equation}
L \left(\mathbf{p} \; | \; \mathbf{f}^{v,c} \right) = g^{v,c} \cdot \prod_n f^{v,c}_{\rho_n},
\label{eqn:likelihoodDP}
\end{equation}
Then, the prediction $\tilde{c} \in \{0,1\}$ of the class for $\mathbf{p}$ is determined by application of the Maximum a Posteriori (MAP) decision rule. But recall that there is a bias towards $c = 1$ for each course (see Fig. \ref{fig:basicInfo}). As a result, we introduce a term $b_v \geq 0$ into MAP, which will be tuned through the cross validation procedure described in Sec. \ref{ssec:procedure}:
\begin{equation}
\tilde{c} = \begin{cases} 1 & g^{v,1} L \left(\mathbf{p} \; | \; \mathbf{f}^{v,1} \right) > g^{v,0} L \left(\mathbf{p} \; | \; \mathbf{f}^{v,0} \right) + b_v \\ 0 & g^{v,1} L \left(\mathbf{p} \; | \; \mathbf{f}^{v,1} \right) < g^{v,0} L \left(\mathbf{p} \; | \; \mathbf{f}^{v,0} \right) + b_v \\ \mathbb{I}_{\{U \geq g^{v,0} \}} & \mbox{otherwise} \end{cases},
\label{eqn:maxlikelihood}
\end{equation}
where $g^{v,c} = |\mathcal{U}_T^{v,c}| / |\mathcal{U}_T^{v}|$ is the estimated class bias for video $v$, and $U$ denotes a random number drawn from $[0,1]$.

\begin{figure*}[t]
\vspace{-0.2in}
\centering
\subfloat[`FMB']{
\includegraphics[scale=0.29]{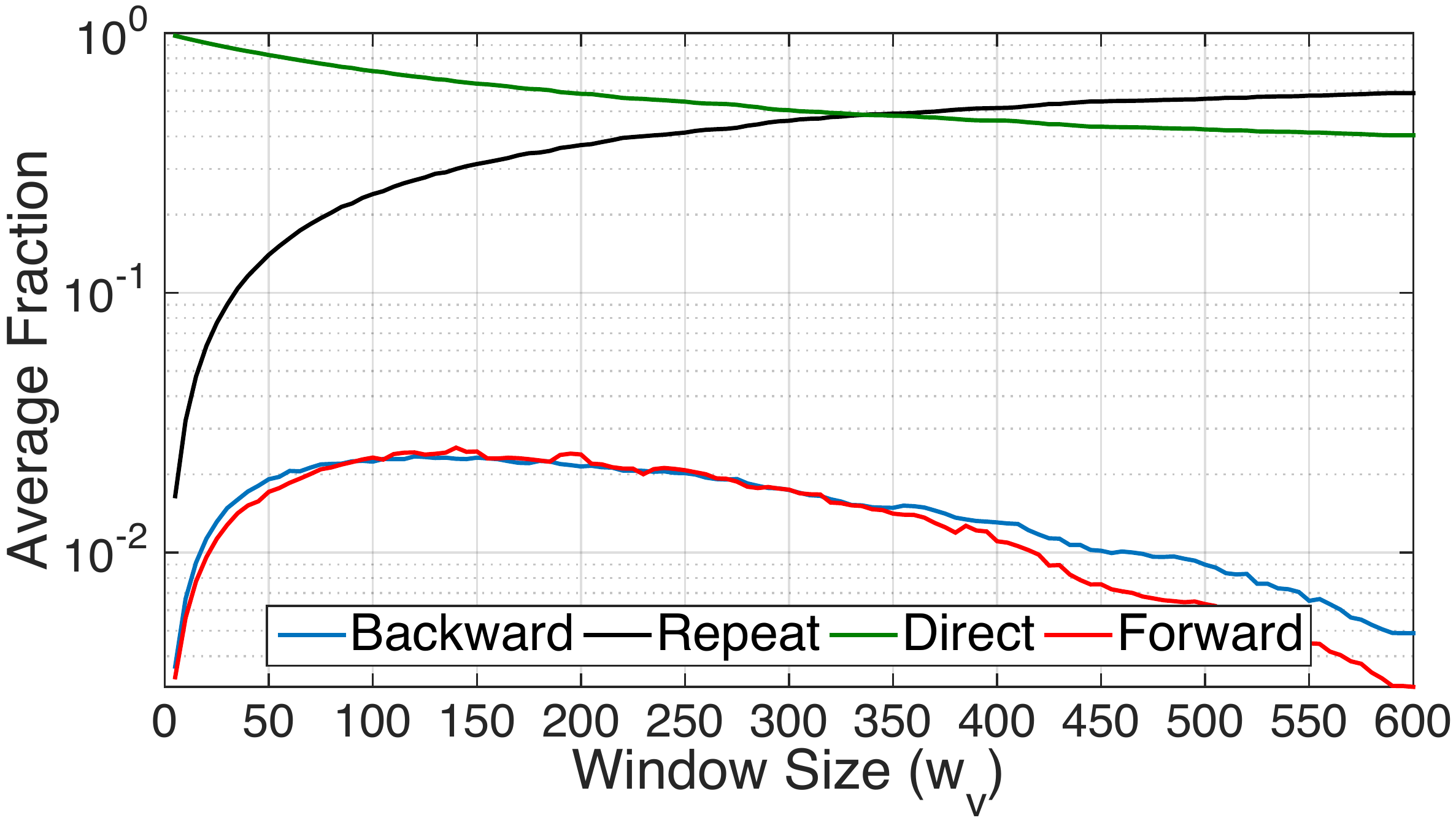}} \qquad \qquad \qquad
\subfloat[`NI']{
\includegraphics[scale=0.29]{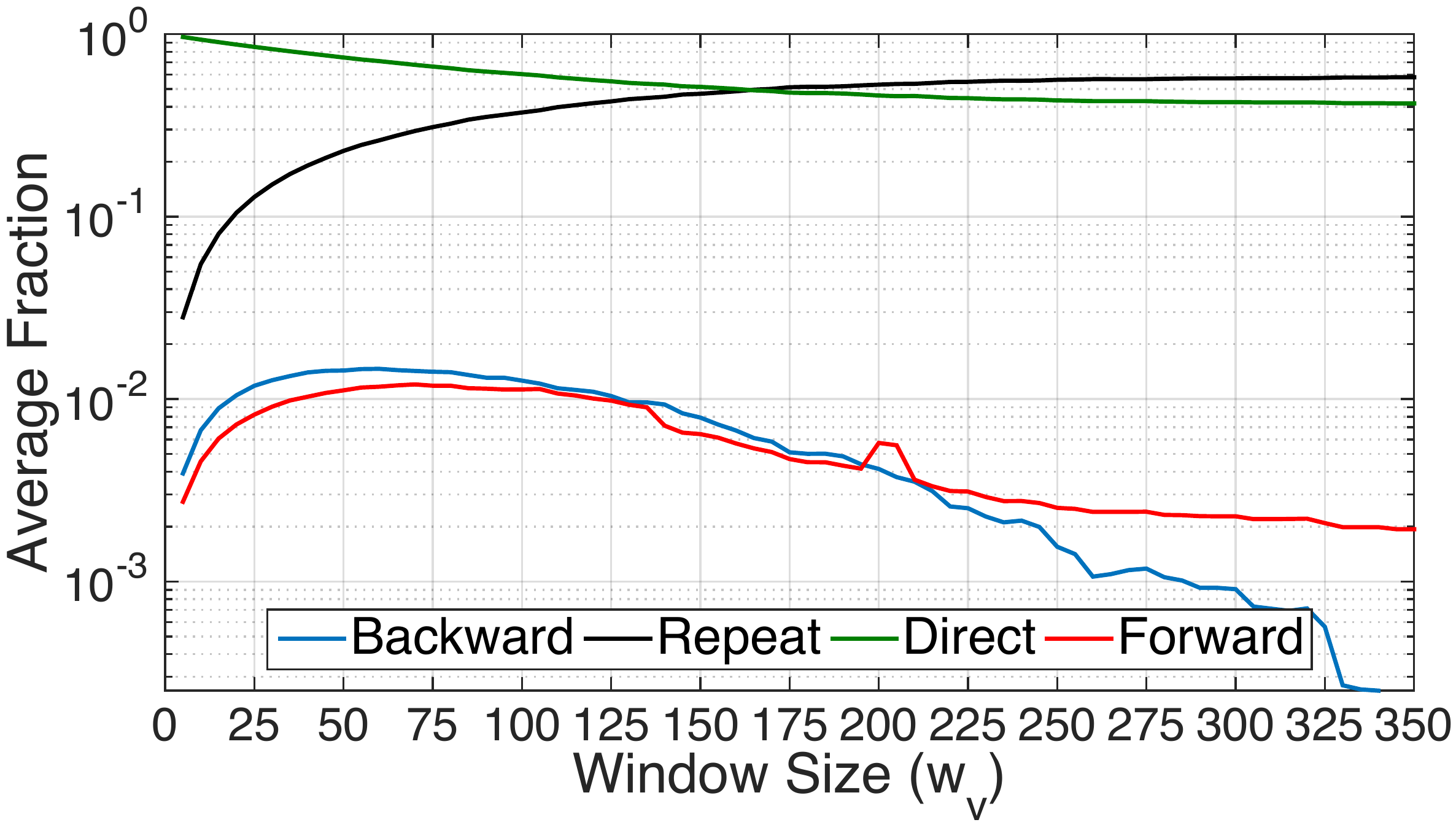}}
\vspace{-0.05in}
\caption{Plot of the fraction of local (repeat and direct) and non-local (backward and forward) transitions for each window size $w_v$, averaged over all UV Pairs for each position and video $v$, for each dataset. Clearly, the fraction of non-local transitions is very low in each case, reaching a maximum of 2.4\% for forward transitions in `FMB' (at $w_v = 150$).}
\label{fig:trans}
\vspace{-0.25in}
\end{figure*}

\subsection{Transition-Based Modeling}
\label{ssec:transModel}
In modeling transitions between positions, we will only consider one-step transitions. This is common in webpage clickstream analysis (\eg \cite{wang2013you}), and will be useful here since the state spaces we consider can be large, depending on $w_v$.\footnote{This may not be ideal because unlike sequences of webpages, learning builds on itself. It is harder to estimate higher order transitions due to position-specific data sparsity. We still see substantial benefit with a one-step model.}

\subsubsection{Aggregating non-local transitions}
The cohort estimator for a Markov Chain model uses the fraction of transitions from state $i$ to $j$ in estimating the probability of transitioning from $i$ to $j$ \cite{norris1998markov}. We found this model not appropriate here, because the number of transitions between two non-local positions is rather sparse, implying that there is not enough data to estimate these specific transitions.

To see this, we inspect the sequences $\mathbf{p}^{v,c}$ for varying $w_v$. In particular, for each position in video $v$, we first find the total number of times each type of transition from Sec. \ref{sssec:modelFactors} occurs, aggregated across the UV pairs. Then, we sum these totals over all positions, and find the fraction of each type of transition. We repeat this for each $w_v \in \{5,10,...,600\}$ (\ie through 10 min), and then average across the videos $v$ for each $w_v$. Fig. \ref{fig:trans} shows the result for each course, from which we make two observations for local and non-local transitions:

\para{(i) Tradeoff between local transitions}: As $w_v$ increases, the percentage of repeat transitions increases monotonically (1.7\% to 59\% for FMB, 2.8\% to 58\% for NI), while the percentage of direct transitions decreases monotonically (98\% to 40\% for FMB, 97\% to 41\% for NI). This is to be expected, since each position is increasing in size with $w_v$.

\para{(ii) Infrequency of non-local transitions}: The vast majority of transitions are local. For example, from Fig. \ref{fig:trans}, the largest fraction of backward transitions is 2.3\% in `FMB', at $w_v = 120$.

As a result of the second observation, the models that follow will aggregate all observed forward transitions to form a single, uniform probability at each position, and likewise for backward transitions. To this end, we define $\mathcal{I}_{i,k} = \{1,...,i-1\} \mbox{ for } k = 1$; $\{i\} \mbox{ for } k = 2$; $\{i + 1\} \mbox{ for } k = 3$; and $\{i+2,...\} \mbox{ for } k = 4$ to be the set of states constituting a backward ($k = 1$), repeat ($k = 2$), direct ($k = 3$), and forward ($k = 4$) transition at position $i$.

\begin{figure*}[t]
\vspace{-0.2in}
\footnotesize
\centering
\subfloat[`FMB']{
\begin{tabular}{c|cc|cc|cc|cc}
     & \multicolumn{2}{c}{$w_v$} & \multicolumn{2}{c}{$b_v$} & \multicolumn{2}{c}{Acc} & \multicolumn{2}{c}{F1} \\
     & avg & s.d. & avg & s.d. & avg & s.d. & avg & s.d. \\ \hline
SR & - & - & - & - & 0.510 & 0.073 & 0.573 & 0.109 \\
DP & 176 & 116 & 4.9E-5 & 1.3E-4 & 0.569 & 0.080 & 0.645 & 0.132 \\
DT & 263 & 109 & 3.5E-5 & 1.0E-4 & 0.572 & 0.084 & 0.614 & 0.165 \\
CT & 212 & 99 & 2.1E-6 & 3.7E-6 & 0.558 & 0.085 & 0.619 & 0.162 \\
\end{tabular}}
\subfloat[`NI']{
\begin{tabular}{c|cc|cc|cc|cc}
     & \multicolumn{2}{c}{$w_v$} & \multicolumn{2}{c}{$b_v$} & \multicolumn{2}{c}{Acc} & \multicolumn{2}{c}{F1} \\
     & avg & s.d. & avg & s.d. & avg & s.d. & avg & s.d. \\ \hline
SR & - & - & - & - & 0.531 & 0.069 & 0.607 & 0.108 \\
DP & 75 & 35 & 3.2E-4 & 7.6E-4 & 0.589 & 0.093 & 0.654 & 0.176 \\
DT & 105 & 72 & 3.7E-3 & 7.8E-3 & 0.587 &  0.099 & 0.652 & 0.152 \\
CT & 71 & 38 & 1.6E-5 & 3.3E-5 & 0.587 & 0.097 & 0.661 & 0.165 \\
\end{tabular}}
\vspace{-0.05in}
\caption{Summary of the tuned parameters window size ($w_v$) and likelihood bias ($b_v$), and of the performance metrics accuracy (Acc) and F1, obtained across the videos for each course. avg and s.d. are calculated over the averages on the 10 evaluation sets for each video.}
\label{fig:modelSummary}
\vspace{-0.2in}
\end{figure*}

\parab{Discrete Time Transitions (DT)}. In this model, we discretize time, discounting the interarrival times. Let $\mathbf{F}^{v,c} = [ f_{i,k} ]^{v,c} \in [0, 1]^{N(w_v),4}$ be the matrix of transition probabilities, where $f^{v,c}_{i,k}$ is the probability that the next position will be in $\mathcal{I}_{i,k}$ given the current is $i$. We also assume that the transitions are homogeneous, \ie independent of time $n$.

Considering the sequences of positions $\mathbf{p}$ across users $u \in \mathcal{U}^{v,c}_T$, we obtain the number transitions from $i$ to $k$ as
\begin{equation}
O^{v,c}_{i,k} = \sum_{u \in \mathcal{U}^{v,c}_T} \sum_{n} \mathbb{I}_{\{ \rho_n = i, \; \rho_{n+1} \in \mathcal{I}_{i,k} \}}.
\label{eqn:numTrans}
\end{equation}
From (\ref{eqn:numTrans}), we estimate $f^{v,c}_{i,k} = O^{v,c}_{i,k} / \sum_{j} O^{v,c}_{i,j}$, and the likelihood of $\mathbf{p}$ from user $u \in \mathcal{U}^v_E$ on video $v$ is
\begin{equation}
L \left(\mathbf{p} \; | \; \mathbf{F}^{v,c} \right) = f^{v,c}_{\rho_1} \cdot \prod_{n} f^{v,c}_{\rho_n, \rho_{n+1}},
\label{eqn:likelihoodDis}
\end{equation}
where $f^{v,c}_{p_1}$ is the distribution at the initial position $\rho_1$ of $\mathbf{p}$, obtained from (\ref{eqn:cohortDP}). The MAP for DT is the same as in (\ref{eqn:maxlikelihood}), except with (\ref{eqn:likelihoodDis}) in place of (\ref{eqn:likelihoodDP}).

\parab{Continuous Time Transitions (CT)}. This model incorporates the interarrival times between transitions. Rather than computing the time-varying transition probabilities, we instead work with the transition rates \cite{norris1998markov}. To this end, we define $\mathbf{Q}^{v,c} = [ q_{i,k} ]^{v,c} \in \mathcal{R}^{N(w_v),4}$ as the transition rate matrix for the model, where $q_{i,k}$, $k \neq 2$ represents the rate of departure from position $i$ and arrival at a position in $\mathcal{I}_{i,k}$.

Let $\mathbf{r}^{v,c} = [r_i]^{v,c} \in \mathcal{R}^{N(w_v)}$ be the vector of the total time spent by $\mathcal{U}_T^{v,c}$ in state $i$. These terms are estimated as
\begin{equation}
r^{v,c}_i =\sum_{u \in \mathcal{U}^{v,c}_T} \sum_{n} \mathbb{I}_{\{\rho_n = i\}} \cdot d_n ,
\label{eqn:holdingTime}
\end{equation}
where $d_n$ is the duration of event $n$ in $\mathbf{p}$ (see Sec. \ref{ssec:evSpec}). In estimating the $q_{i,k}$, we must also obtain the number of transitions from $i$ to $k$ over users $u \in \mathcal{U}^v_T$, \ie the $O^{v,c}_{i,k}$ from (\ref{eqn:numTrans}); with this, the $q^{v,c}_{i,k}$ terms are estimated as
\begin{equation}
q^{v,c}_{i,k} = \begin{cases} O^{v,c}_{i,k} / r^{v,c}_i & \;\; k \neq 2 \\ - \sum_{k \neq 2} q^{v,c}_{i,k} & \;\; k = 2 \end{cases},
\label{eqn:qEst}
\end{equation}
Finally, the likelihood of sequence $\mathbf{p}$ for $u \in \Omega^{v}_E$ is computed:
\begin{equation}
L \left( \mathbf{p} \; | \; \mathbf{Q}^{v,c} \right) = \prod_{i,k; k \neq 2} (q^{v,c}_{i,k})^{o_{i,k}} \exp \left( -q^{v,c}_{i,k}  \cdot T_i \right),
\label{eqn:likelihoodCon}
\end{equation}
where $o_{i,k} = \sum_n \mathbb{I}_{\{\rho_n = i, \; \rho_{n+1}  \in \mathcal{I}_{ik} \}}$, $k \neq 2$ is the number of transitions from $i$ to $k$ for the sequence $\mathbf{p}$, and $T_i = \sum_n \mathbb{I}_{\{ \rho_n = i \}} \cdot d_n$ is the time spent by $\mathbf{p}$ in $i$. Once again, MAP is as in (\ref{eqn:maxlikelihood}), except with (\ref{eqn:likelihoodCon}) in place of (\ref{eqn:likelihoodDP}).

We also considered another position-based model, Continuous Time Positions (CP), which used the time spent at each position in likelihood computation. We omit it because its results were strictly lower than the other three models.

\section{Prediction Evaluation}
\label{sec:modelEval}
In this section, we evaluate the performance of the models described in Section \ref{sec:positions}. We pose the following questions:

\parab{1}. How beneficial is it to include positions and transitions for CFA prediction on individual videos?

\parab{2}. Is one of position or transition-based model clearly better than the other, or would some combination be the best?

\parab{3}. Is it beneficial to include position durations? \\

\parab{Skewed-Random} (SKR). We will also consider an algorithm that does not make use of clickstream data, to act as a baseline for evaluating the gain from incorporating behavior. SKR finds the CFA bias $g^{v,1}$ over the training set $\mathcal{U}_T^v$, and predicts $c = 1$ $g^{v,1}$ of the time (similar to the baseline used in \cite{brinton2015prediction}). Note that in our application of CFA prediction for individual videos, more sophisticated baselines that would leverage similarities across users and/or quizzes without behavioral data (\eg collaborative filtering like in \cite{toscher2010collaborative,bergner2012model}) are not applicable.

\subsection{Procedure}
\label{ssec:procedure}
\parab{Metrics}. Let TP, FP, TN, and FN be the number of true and false positives, and true and false negatives obtained by a model on an evaluation set. The first metric we consider is \textit{accuracy}, \ie $(\mbox{TP} + \mbox{TN})/(\mbox{TP} + \mbox{FP} + \mbox{TN} + \mbox{FN})$. Since the quizzes are biased towards CFA (see Fig. \ref{fig:evtStats}), we found that unconstrained maximization of accuracy during the tuning procedure (described below) led to high recall (rec), \ie $\mbox{TP} / (\mbox{TP} + \mbox{FN})$ but low precision (prec), \ie $\mbox{TP} / (\mbox{TP} + \mbox{FP})$. To avoid this, we will subject tuning to the constraint that the chosen parameters have at least 25\% of the truly negative samples predicted negative, and likewise for the positives. To this end, the second metric we consider is the standard (balanced) \textit{F1 score}, obtained as $2 \cdot (\mbox{prec} + \mbox{rec}) / (\mbox{prec} + \mbox{rec})$ \cite{murphy2012machine}. As the harmonic mean of precision and recall, F1 is limited by the minimum of the two, capturing the tradeoff between them that is induced by this constraint.

\begin{figure*}[t]
\vspace{-0.1in}
\centering
\subfloat[`FMB', accuracy]{
\includegraphics[scale=0.38]{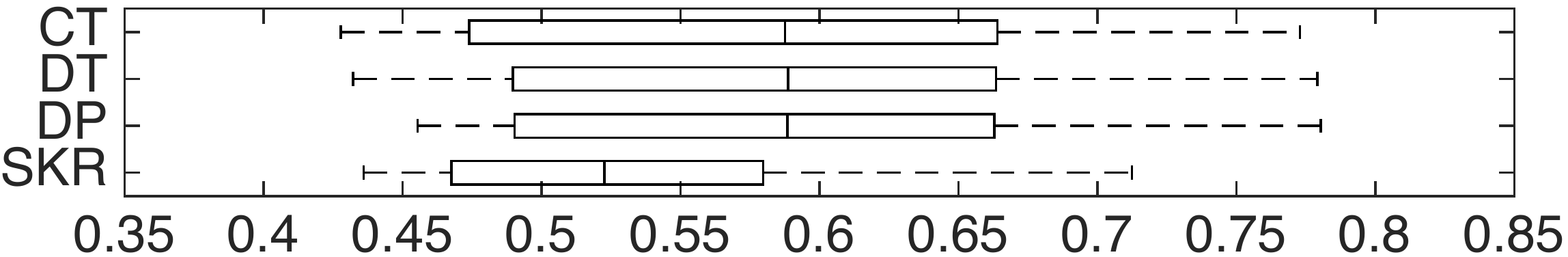}}
\subfloat[`FMB', F1]{
\includegraphics[scale=0.38]{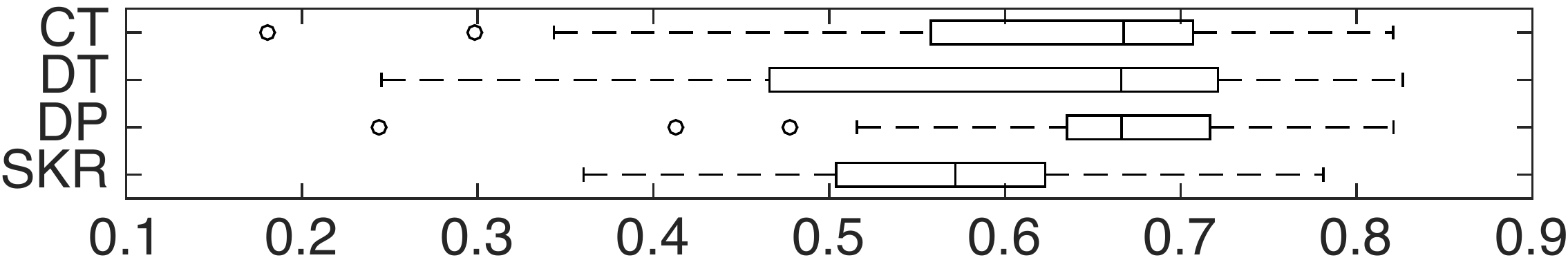}} \\
\vspace{-0.1in}
\subfloat[`NI', accuracy]{
\includegraphics[scale=0.38]{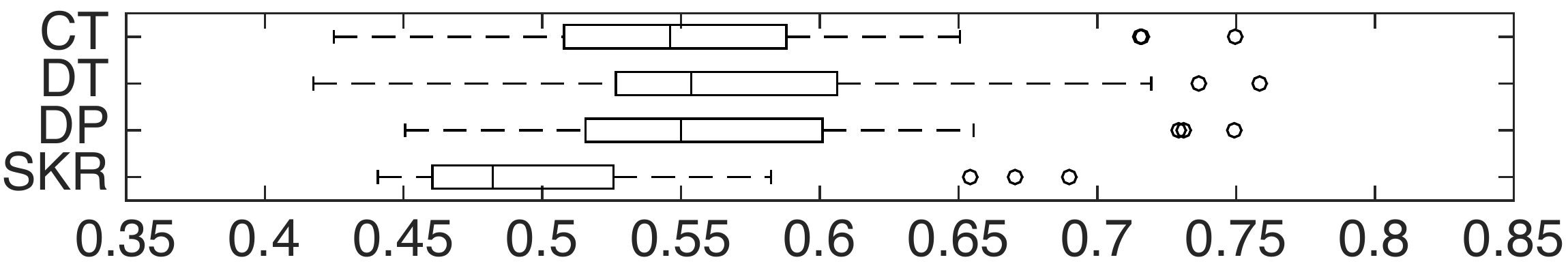}}
\subfloat[`NI', F1]{
\includegraphics[scale=0.38]{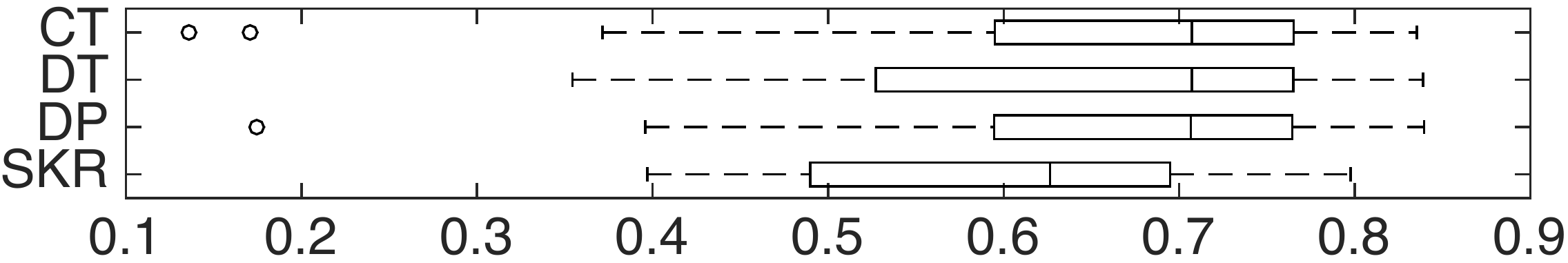}}
\vspace{-0.1in}
\caption{Boxplots of CFA prediction performance across both courses, considering accuracy and F1. Here, each datapoint is the obtained performance on one of the videos considered. Overall, we see that DP, DT, and CT outperform SKR for both metrics, and especially for accuracy, while CP performs comparable to SKR.}
\label{fig:accF1Box}
\vspace{-0.2in}
\end{figure*}
\begin{figure*}
\centering
\subfloat[`FMB', accuracy]{
\small
\begin{tabular}{c| C{13mm} C{13mm} C{13mm} C{13mm}}
 & SKR & DP & DT & CT \\ \hline
SKR & -- & 2.5E-3$^{**}$ & 2.2E-3$^{**}$ & 0.018$^{*}$ \\
DP & 2.5E-3$^{**}$ & -- & 0.75 & 0.72 \\
DT & 2.2E-3$^{**}$ & 0.75 & -- & 0.28 \\
CT & 0.018$^{*}$ & 0.72 & 0.28 & -- \\
\end{tabular}} \qquad
\subfloat[`FMB', F1]{
\small
\begin{tabular}{c| C{13mm} C{13mm} C{13mm} C{13mm}}
 & SKR & DP & DT & CT \\ \hline
SKR & -- & 0.014$^{*}$ & 0.16 & 0.065  \\
DP & 0.014$^{*}$ & -- & 0.85 & 0.77 \\
DT & 0.16 & 0.85 & -- & 0.98 \\
CT & 0.065 & 0.77 & 0.98 & -- \\
\end{tabular}} \\
\subfloat[`NI', accuracy]{
\small
\begin{tabular}{c| C{13mm} C{13mm} C{13mm} C{13mm}}
 & SKR & DP & DT & CT \\ \hline
SKR & -- & 8.0E-3$^{**}$ & 0.019$^{*}$ & 0.015$^{*}$ \\
DP & 8.0E-3$^{**}$ & -- & 0.91 & 0.79 \\
DT & 0.019$^{*}$ & 0.91 & -- & 0.94  \\
CT & 0.015$^{*}$ & 0.79 & 0.94 & -- \\
\end{tabular}} \qquad
\subfloat[`NI', F1]{
\small
\begin{tabular}{c| C{13mm} C{13mm} C{13mm} C{13mm}}
 & SKR & DP & DT & CT \\ \hline
SKR & -- & 0.012$^{*}$ & 0.045$^{*}$ & 6.3E-3$^{**}$ \\
DP & 0.012$^{*}$ & -- & 0.90 & 0.99 \\
DT & 0.045$^{*}$ & 0.90 & -- & 0.86 \\
CT & 6.3E-3$^{**}$ & 0.99 & 0.86 & -- \\
\end{tabular}}
\vspace{-0.05in}
\caption{$p$-values ($p$) from applying pairwise WRS tests to the boxplots in Fig. \ref{fig:accF1Box}. A $^{*}$ indicates significance at $p \leq 0.05$, and $^{**}$ at $p \leq 0.01$.}
\label{fig:resWRS}
\vspace{-0.2in}
\end{figure*}

Even a few percent improvement in these metrics can be a substantial gain for CFA prediction. As a reference, in KDD Cup 2010 for CFA prediction there was only 1\% improvement from the 132nd to the best score on the leaderboard \cite{brinton2015prediction}.

\parab{Training and testing}. We consider $N$ evaluation iterations for each video. In each iteration, we use the following procedure:
\begin{enumerate}
\item Divide the elements of $\mathcal{U}^v$ into $K$ disjoint folds $\mathcal{U}^v_1, \mathcal{U}^v_2, ..., \mathcal{U}^v_K$. In doing so, we randomly allocate samples of CFA and non-CFA to folds, ensuring that the number of class instances is equal across folds (\eg $|\mathcal{U}^{v,c}_k| = |\mathcal{U}^{v,c}_l| \; \forall k,l$).
\item Set $\mathcal{U}^v_E = \mathcal{U}^v_K$ and $\mathcal{U}^v_T = \mathcal{U}^v \setminus \mathcal{U}^v_K$.
\item Using $\mathcal{U}^v_T$, tune the algorithm parameters $w_v$ and $b_v$ through the parameter tuning procedure described below.
\item With the tuned values, train the quantities required to compute the likelihoods and MAP of each model over the full $\mathcal{U}^v_T$, and evaluate on $\mathcal{U}^v_E$.
\end{enumerate}

The results for each metric are averaged over the $N$ iterations. In our evaluation, we set $N = 10$ and $K = 5$.

\parab{Parameter tuning}. Each algorithm has two parameters that must be tuned: the video width $w_v$ and the likelihood bias $b_v$. To do this, we apply Cross-Validation (CV) as described in \cite{murphy2012machine} over the $K-1$ training set partitions. The following is the procedure for each CV iteration $k \in \{1,...,K-1\}$:

\begin{enumerate}
\item Set $\mathcal{U}^v_C = \mathcal{U}^v_k$ and $\mathcal{U}^v_R = \mathcal{U}^v_T \setminus \mathcal{U}^v_k$.
\item Obtain the results of training on $\mathcal{U}^v_R$ and testing on $\mathcal{U}^v_C$ for each pair $(w_v,b_v) \in \{5,10,...,20,30,45,...,600\} \times \{0,$ $2^{-60}, 2^{-58}, ..., 1\}$, \ie a total of 1,376 pairs.
\end{enumerate}

In the end, we select the combination of parameters $(w_v,b_v)$ which yields the highest average accuracy over the CV iterations, subject to the constraint described with the metrics. Note that for $w_v$, we choose this set since (i) 5 sec corresponds to the threshold of combining repeat events (see Sec. \ref{ssec:evSpec}), and (ii) 600 is close to the minimum video length in both courses. For both parameters, these choices ensured that most selections across videos did not lie on one of the grid endpoints.

\subsection{Results and Discussion}
\label{ssec:results}
Since there is a sharp dropoff in quiz participation over time, we only consider those for which there are at least 100 samples of both CFA and non-CFA instances, so that there at least 20 samples from each group in each of the five folds. This leaves a total of 24 videos for `FMB' and 32 for `NI'.

\parab{Overview of results}. Summary information on the tuned $w_v$ and $b_v$ values, as well as the two performance metrics -- Accuracy (Acc) and F1 -- can be found for each course in Fig. \ref{fig:modelSummary}. Here, we give the average (avg) and standard deviation (s.d.) of these values across videos. The distribution of the performance values are plotted for each course in Fig. \ref{fig:accF1Box}; in each box, the performance on one video is one data point.

From Fig. \ref{fig:accF1Box}, we can see immediately that the DP, DT, and CT algorithms perform substantially better than SKR overall. Further, the improvement is higher for accuracy than for F1, which is expected since the tuning monitors accuracy. In order to test for significance in the performance differences between each pair of models, we run a WRS test (as in Sec. \ref{sec:clickstream}) for the null hypothesis that there is no difference between the distributions in Fig. \ref{fig:accF1Box}. The resulting p-values ($p$) from these tests are tabulated in Fig. \ref{fig:resWRS}, and verify the differences.

\begin{figure*}[t]
\vspace{-0.15in}
\centering
\subfloat[`FMB', Accuracy]{
\includegraphics[scale=0.32]{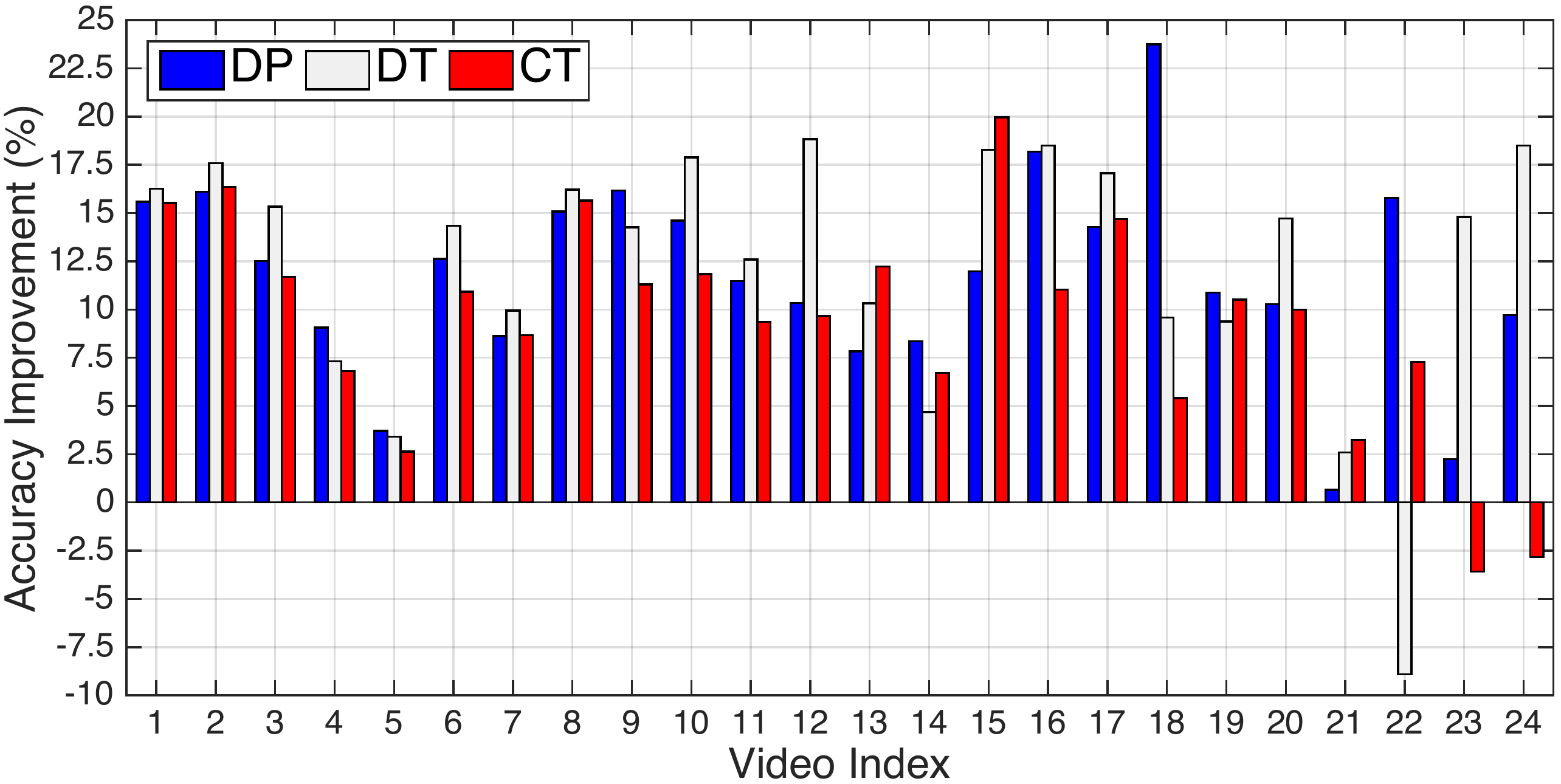}} \hspace{0.2in}
\subfloat[`FMB', F1]{
\includegraphics[scale=0.32]{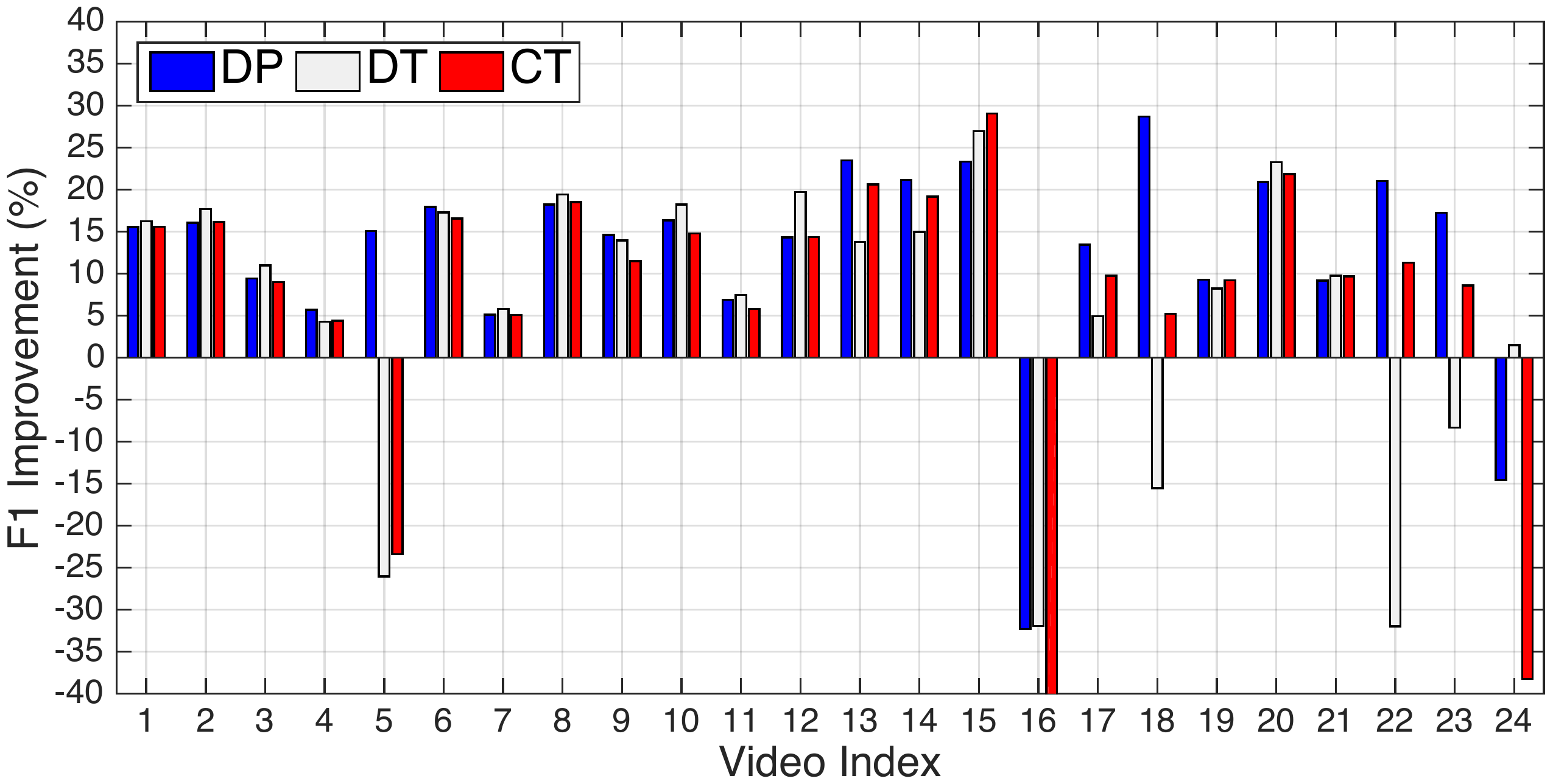}} \\
\vspace{-0.1in}
\subfloat[`NI', Accuracy]{
\includegraphics[scale=0.32]{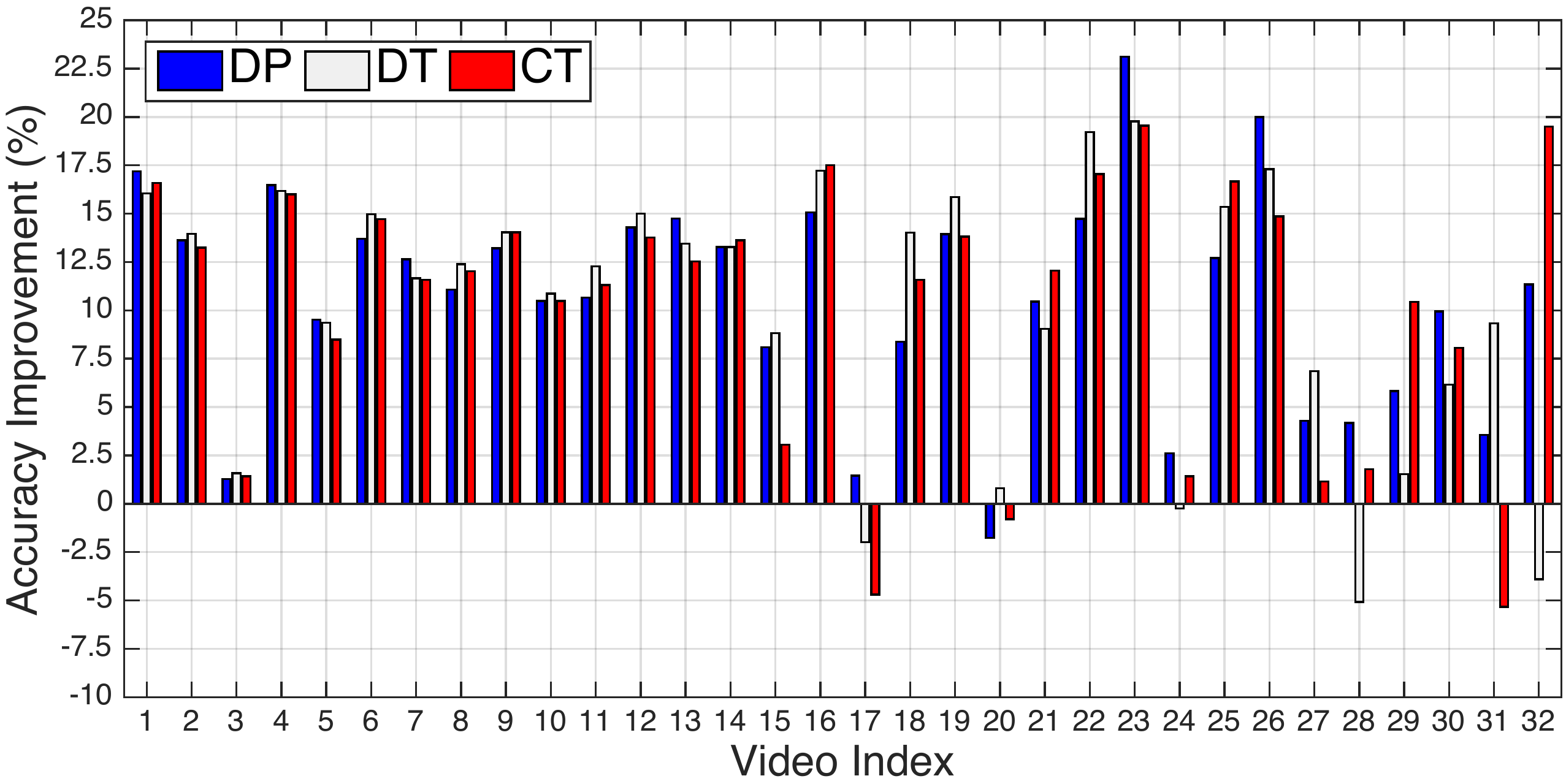}} \hspace{0.2in}
\subfloat[`NI', F1]{
\includegraphics[scale=0.32]{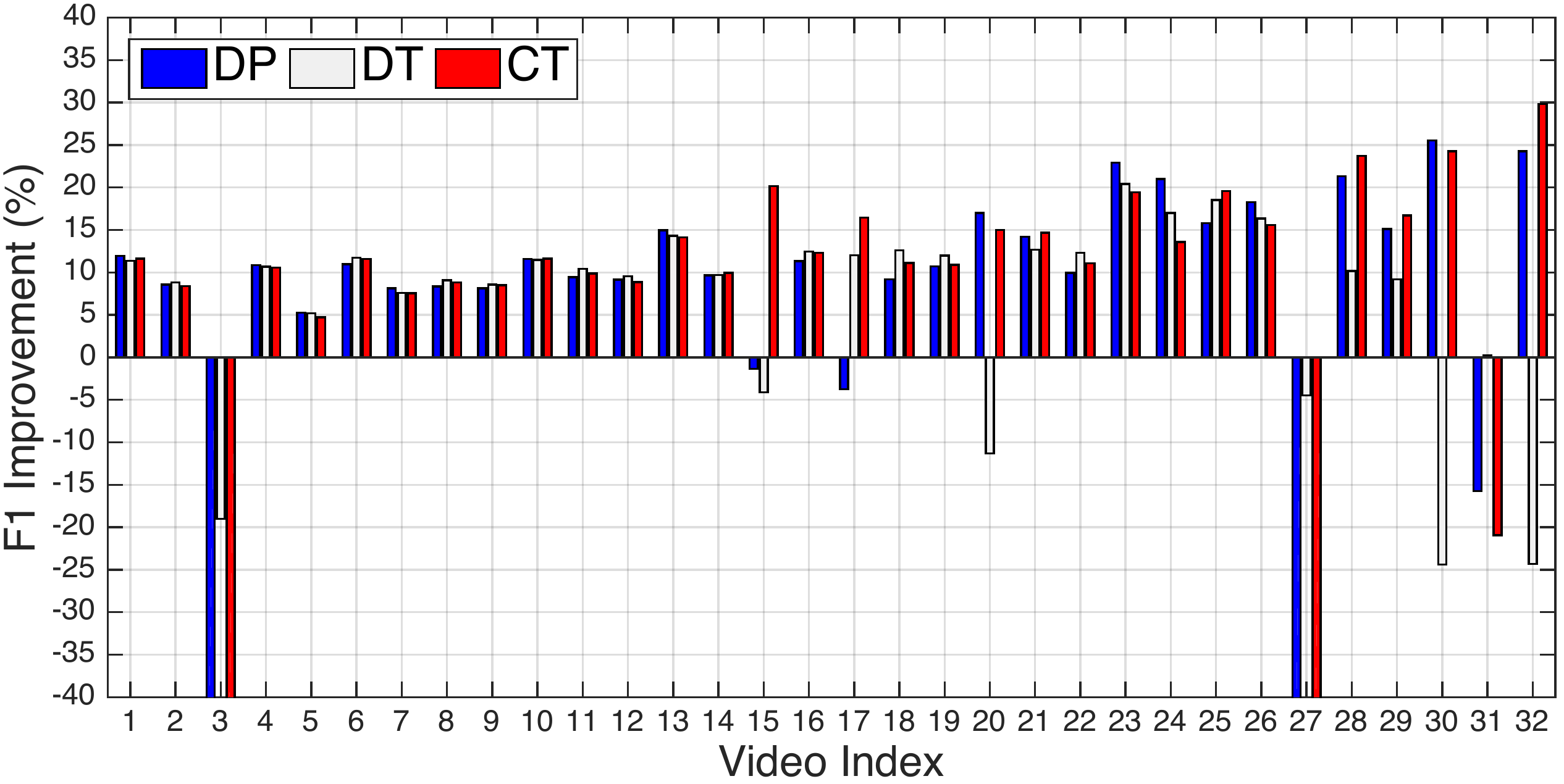}}
\caption{Percent improvements of each algorithm relative to SKR for individual videos, in each course for each metric. Consistent with Fig. \ref{fig:accF1Box}, we see that each algorithm outperforms SKR in the vast majority of cases, except for six videos with respect to F1.}
\label{fig:accF1Bar}
\vspace{-0.2in}
\end{figure*}

Finally, in Fig. \ref{fig:accF1Bar}, we plot the percent increase in performance for each of the algorithms relative to SKR on each of the videos, for a more specific case-by-case comparison.

\parab{1: Benefit of clickstream data}. We assess how beneficial the position and transition information is for prediction by comparing each of the algorithms to SKR.

\para{Accuracy}: Considering accuracy first, refer to Fig. \ref{fig:accF1Box}(a\&c). Here, we see that the DP, DT, and CT models are clearly shifted to the right relative to SKR, indicating higher quality. For `FMB', the shift in the mean of DP and DT relative to SKR is roughly 12\%, and of CT is 9\%; for `NI', the improvements are roughly 11\% for each of the algorithms. From Fig. \ref{fig:resWRS}, we see that this difference is also statistically significant for each algorithm across both courses, with $p < 0.02$ in each case.

As for individual videos in Fig. \ref{fig:accF1Bar}(a\&c), we see that each algorithm outperforms SKR in the vast majority of cases, across both datasets. The fraction of times in which DP, DT, and CT outperform SKR in `FMB' (`NI') is 100\% (97\%), 96\% (88\%), and 92\% (91\%), respectively.

\para{F1 score}: For F1, refer to Fig. \ref{fig:accF1Box}(b\&d). Again, we see that DP, DT, and CT are shifted to the right relative to SKR overall, but not as substantially. This is especially true for DT, which has the highest range of F1 scores. For DP, the increase in mean performance of roughly 13\% for `FMB' and 8\% for `NI' are both significant, with $p < 0.02$ from Fig. \ref{fig:resWRS}. Both DT and CT have significant improvements of 7\% and 9\% in `NI' ($p < 0.05$); the improvements of 7\% and 8\% in `FMB' are also substantial, but not significant ($p > 0.06$). The number of videos in which each algorithm outperforms SKR is also lower than for accuracy; for DP, DT, and CT, these numbers for `FMB' (`NI') are 92\% (84\%), 79\% (81\%), and 88\% (91\%).

In Fig. \ref{fig:accF1Bar}(b\&d), we remark that there are a total of six videos (indexes 5, 16, and 24 in `FMB', and 3, 27, and 31 in `NI') where most of the DP, DT, and CT algorithms perform \textit{substantially worse} than SKR in F1-score. These videos also correspond to the outliers observed below the first quartiles in Fig. \ref{fig:accF1Box}(b\&d). One would expect that these would be instances where SKR already had high performance due to a high bias (skew) in favor of either CFA or non-CFA (\eg a video with an easy or a hard quiz). Surprisingly, the opposite is true: the F1 and accuracy scores obtained by SKR on these five videos are all within the bottom nine of all videos. There is also no consistency among the CFA biases (half above $0.5$, half are below). Further, there are other videos with biases in the same ranges where the algorithms outperform SKR substantially.

\parab{2: Positions vs. transitions}. For this, we compare DP to DT. In terms of accuracy, in Fig. \ref{fig:accF1Box}(a\&c) we see that the algorithms are comparable for both courses on average. As for F1 in Fig. \ref{fig:accF1Box}(b\&d), DP has modestly better average performance, especially for `FMB' where it has an improvement of roughly 5\%. DT has a higher range in each case (excluding outliers), with generally lower performance than DP below quartile Q2 (\eg in F1 for `FMB') but, in accuracy for `FMB', also higher above Q2. When considering individual videos in Fig. \ref{fig:accF1Bar}, DT and DP each perform better in roughly 50\% of the cases, with the exception of accuracy in `FMB' for which DT is higher the majority of the time. Overall, the differences between DT and DP are not statistically significant for either course or metric, with $p \geq 0.75$ in all cases in Fig. \ref{fig:resWRS}.

\parab{3: Discrete vs. continuous}. Finally, we compare DT to CT. In Fig. \ref{fig:accF1Box}, in terms of accuracy: For `FMB', DT is shifted to the right by roughly 3\% relative to CT, whereas for `NI', the algorithms are comparable. As to F1-score, while DT and CT are comparable overall, the distribution for CT is slightly shifted to the right for both courses. Considering individual videos, DT outperforms CT on more videos in each of the four cases in Fig. \ref{fig:accF1Bar}. In particular, for accuracy (F1), it outperforms in 75\% (58\%) of the cases for `FMB', and 69\% (63\%) for `NI'. Still, overall, the differences are not statistically significant for either course or metric, with $p \geq 0.28$ in all cases.

\parab{Key messages}. Many aspects of position-based video behavior are useful for CFA prediction: the frequency of visits to each position (DP), the frequency of transitions between positions (DT), and transitions incorporating holding times (CT). These benefits are also measured on individual videos, which underscores the applicability of these models to situations where there is not a lot of information across multiple lectures, \eg for quick detection early in a course. Both positions and transitions can be useful; DP, DT, and CT are comparable overall, performing better on different sets of videos.

Each of the algorithms tested here employ feature spaces that are representing user behavior directly; namely, positions visited and transitions. Higher quality predictions may be attainable by passing these through more complicated machine learning algorithms (\eg kernel SVM) to learn over higher dimensional spaces. A significant advantage of our feature spaces, though, is their \textit{natural interpretation} in terms of learner actions, which can be related to CFA results. An interesting avenue of future work would be to use the position and transition matrices inferred over the CFA classes to generate recommendations guiding learner behavior in real time.

\section{Related Work}
\label{sec:related}
We discuss recent, key works on MOOC, student video-watching analysis, and CFA prediction.

\parab{MOOC studies}. With the proliferation of MOOC in recent years, there have been a number of analytical studies on these platforms. Some have focused on a more general analysis of all learning modes, \eg \cite{anderson2014engaging,kizilcec2013deconstructing} studied learner engagement variation over time and across courses. Others have focused on specific modes, \eg in terms of forums, \cite{brinton2014learning} analyzed the decline in participation over 73 courses. Our work is fundamentally different from these works in that is explores the \textit{association} between behavior with two modes: video and assessment.

\parab{Video-watching analysis}. Most existing work on learner video-watching behavior \cite{kim2014understanding,brinton2015prediction,aiken2014student} has focused session-level user characteristics (\eg rewatching sessions), rather than click-level information. The work in \cite{sinha2014your} is most similar to ours, since it is also concerned with recurring patterns in clickstream sequences for MOOC users. The authors define a mapping of subsequences of events to predefined behavioral actions (\eg skipping, slow watching) and perform approximate string search to locate these behaviors in clickstreams. Our work on motif identification differs in two important ways: (i) rather than assuming a predefined set of actions, we extract the recurring sequences directly using motif identification algorithms, and (ii) we are concerned with mapping motifs to efficacy, in contrast to \cite{sinha2014your} where the objective is to predict engagement, next click, and dropout.

\parab{Performance prediction}. Researchers have developed predictors for whether a student will be CFA or not on a question in traditional education settings. Collaborative filtering algorithms have been applied as classification models for this purpose (\eg \cite{toscher2010collaborative,bergner2012model}). Others have probabilistic graphical models (PGMs) \cite{pardos2011accepted}, when there is coarse-granular information collected (\eg course difficulty) over multiple sessions. Recently, \cite{lan2014time} developed SPARFA-Trace, which traces a learner's knowledge through the sequence of material accessed and questions answered. Compared with these works, ours is unique in that (i) it focuses on relating click-level data -- video-watching behavior -- to performance, and (ii) it focuses on prediction within single videos. The recent work of \cite{brinton2015prediction} studied the predictive capability of session-level video-watching quantities computed from clickstream data (\eg the fraction of the video watched and the number of rewinds), considering multiple users and videos in the course simultaneously. Focusing on individual videos, our models are instead position-dependent, and the improvements in accuracy relative to the baseline that we obtain are strictly higher than those cited here (3\% increase to the same baseline). Overall, we emphasize that the models used in each of these other works are not readily applicable to our setting, because we focus on the case of individual videos where similarities among users/quizzes is not available.

\parab{Webpage clickstream analysis}. Webpage clickstream analysis \cite{wang2013you,gunduz2003web,speiser2012nested} remains an active area of research. Video-watching clickstreams are fundamentally different than these applications, which concern transitions between webpages rather than behavior within a single window.

\section{Conclusion and Future Work}
In this work, we studied student video-watching behavior, performance, and their association in MOOC. In doing so, we formalized two frameworks for representing user clickstreams: one based on sequences of events with discretized lengths, and one based on sequences of positions visited. With datasets from two MOOCs encoded in these frameworks, we accomplished two goals: (i) we mined the sequences to identify recurring motifs in user behavior, and discovered that some of these characteristics are significantly associated with CFA and non-CFA quiz submissions; (ii) we proposed models for relating user clickstreams to knowledge gained, and showed how multiple aspects of this behavior can improve CFA prediction quality on individual videos.

There are a number of next steps we are investigating, \eg to use the identified motifs for user and content analytics; to optimize the selection of quantiles used divide the event lengths; to consider position transition durations under a non-exponential assumption; and to see whether prediction improvement can be obtained through higher order transitions.

\ifCLASSOPTIONcompsoc
  \section*{Acknowledgments}
\else
  \section*{Acknowledgment}
\fi
This work was in part supported by ARO grants W911NF-11-1-0036 and W911NF-14-1-0190.

\bibliographystyle{IEEEtran}
{\small \bibliography{references}}

\begin{thebibliography}{10}
\providecommand{\url}[1]{#1}
\csname url@samestyle\endcsname
\providecommand{\newblock}{\relax}
\providecommand{\bibinfo}[2]{#2}
\providecommand{\BIBentrySTDinterwordspacing}{\spaceskip=0pt\relax}
\providecommand{\BIBentryALTinterwordstretchfactor}{4}
\providecommand{\BIBentryALTinterwordspacing}{\spaceskip=\fontdimen2\font plus
\BIBentryALTinterwordstretchfactor\fontdimen3\font minus
  \fontdimen4\font\relax}
\providecommand{\BIBforeignlanguage}[2]{{%
\expandafter\ifx\csname l@#1\endcsname\relax
\typeout{** WARNING: IEEEtran.bst: No hyphenation pattern has been}%
\typeout{** loaded for the language `#1'. Using the pattern for}%
\typeout{** the default language instead.}%
\else
\language=\csname l@#1\endcsname
\fi
#2}}
\providecommand{\BIBdecl}{\relax}
\BIBdecl

\bibitem{brinton2014social}
C.~G. Brinton and M.~Chiang, ``{Social Learning Networks: A Brief Survey},'' in
  \emph{CISS}.\hskip 1em plus 0.5em minus 0.4em\relax IEEE, 2014, pp. 1--6.

\bibitem{brinton2015individualization}
C.~G. Brinton, R.~Rill, S.~Ha, M.~Chiang, R.~Smith, and W.~Ju,
  ``{Individualization for Education at Scale: MIIC Design and Preliminary
  Evaluation},'' \emph{Transactions on Learning Technologies}, vol.~8, no.~1,
  2015.

\bibitem{brinton2014learning}
C.~G. Brinton, M.~Chiang, S.~Jain, H.~Lam, Z.~Liu, and F.~M.~F. Wong,
  ``{Learning About Social Learning in MOOCs: From Statistical Analysis to
  Generative Model},'' \emph{Transactions on Learning Technologies}, vol.~7,
  pp. 346--359, 2014.

\bibitem{anderson2014engaging}
A.~Anderson, D.~Huttenlocher, J.~Kleinberg, and J.~Leskovec, ``{Engaging with
  Massive Online Courses},'' in \emph{WWW}.\hskip 1em plus 0.5em minus
  0.4em\relax ACM, 2014, pp. 687--698.

\bibitem{brinton2015prediction}
C.~G. Brinton and M.~Chiang, ``{MOOC Performance Prediction via Clickstream
  Data and Social Learning Networks},'' in \emph{INFOCOM}.\hskip 1em plus 0.5em
  minus 0.4em\relax IEEE, 2015.

\bibitem{yang2013turn}
D.~Yang, T.~Sinha, D.~Adamson, and C.~P. Rose, ``{Turn on, Tune in, Drop out:
  Anticipating Student Dropouts in Massive Open Online Courses},'' in
  \emph{NIPS Data-Driven Education Workshop}, vol.~11, 2013, p.~14.

\bibitem{sinha2014your}
T.~Sinha, P.~Jermann, N.~Li, and P.~Dillenbourg, ``{Your Click Decides your
  Fate: Inferring Information Processing and Attrition Behavior from MOOC Video
  Clickstream Interactions},'' in \emph{EMNLP}, 2014.

\bibitem{piech2013tuned}
C.~Piech, J.~Huang, Z.~Chen, C.~Do, A.~Ng, and D.~Koller, ``{Tuned Models of
  Peer Assessment in MOOCs},'' in \emph{EDM}, 2013.

\bibitem{kim2014understanding}
J.~Kim, P.~J. Guo, D.~T. Seaton, P.~Mitros, K.~Z. Gajos, and R.~C. Miller,
  ``{Understanding In-Video Dropouts and Interaction Peaks in Online Lecture
  Videos},'' in \emph{L@S}.\hskip 1em plus 0.5em minus 0.4em\relax ACM, 2014,
  pp. 31--40.

\bibitem{stephens2014monitoring}
K.~Stephens-Martinez, M.~A. Hearst, and A.~Fox, ``{Monitoring MOOCs: Which
  Information Sources do Instructors Value?}'' in \emph{L@S}.\hskip 1em plus
  0.5em minus 0.4em\relax ACM, 2014, pp. 79--88.

\bibitem{lan2014time}
A.~S. Lan, C.~Studer, and R.~G. Baraniuk, ``{Time-Varying Learning and Content
  Analytics via Sparse Factor Analysis},'' in \emph{KDD}.\hskip 1em plus 0.5em
  minus 0.4em\relax ACM, 2014, pp. 452--461.

\bibitem{wang2013you}
G.~Wang, T.~Konolige, C.~Wilson, X.~Wang, H.~Zheng, and B.~Y. Zhao, ``{You Are
  How You Click: Clickstream Analysis for Sybil Detection.}'' in \emph{USENIX
  Security}, 2013, pp. 241--256.

\bibitem{sheskin2003handbook}
D.~J. Sheskin, \emph{{Handbook of Parametric and Nonparametric Statistical
  Procedures}}.\hskip 1em plus 0.5em minus 0.4em\relax crc Press, 2003.

\bibitem{bailey2009meme}
T.~L. Bailey, M.~Boden, F.~A. Buske, M.~Frith, C.~E. Grant, L.~Clementi,
  J.~Ren, W.~W. Li, and W.~S. Noble, ``{MEME Suite: Tools for Motif Discovery
  and Searching},'' \emph{Nucleic Acids Research}, p. gkp335, 2009.

\bibitem{norris1998markov}
J.~R. Norris, \emph{{Markov Chains}}.\hskip 1em plus 0.5em minus 0.4em\relax
  Cambridge university press, 1998, no.~2.

\bibitem{toscher2010collaborative}
A.~Toscher and M.~Jahrer, ``{Collaborative Filtering Applied to Educational
  Data Mining},'' \emph{KDD Cup}, 2010.

\bibitem{bergner2012model}
Y.~Bergner, S.~Droschler, G.~Kortemeyer, S.~Rayyan, D.~Seaton, and D.~E.
  Pritchard, ``{Model-Based Collaborative Filtering Analysis of Student
  Response Data: Machine-Learning Item Response Theory},'' in \emph{EDM}.\hskip
  1em plus 0.5em minus 0.4em\relax ERIC, 2012, pp. 95--102.

\bibitem{murphy2012machine}
K.~P. Murphy, \emph{{Machine Learning: A Probabilistic Perspective}}.\hskip 1em
  plus 0.5em minus 0.4em\relax MIT press, 2012.

\bibitem{kizilcec2013deconstructing}
R.~F. Kizilcec, C.~Piech, and E.~Schneider, ``{Deconstructing Disengagement:
  Analyzing Learner Subpopulations in Massive Open Online Courses},'' in
  \emph{LAK}.\hskip 1em plus 0.5em minus 0.4em\relax ACM, 2013, pp. 170--179.

\bibitem{aiken2014student}
J.~M. Aiken, S.-Y. Lin, S.~S. Douglas, E.~F. Greco, B.~D. Thoms, M.~D.
  Caballero, and M.~F. Schatz, ``{Student Use of a Single Lecture Video in a
  Flipped Introductory Mechanics Course},'' \emph{arXiv:1407.2620}, 2014.

\bibitem{pardos2011accepted}
Z.~Pardos and N.~Heffernan, ``{Using HMMs and Bagged Decision Trees to Leverage
  Rich Features of User and Skill from an Intelligent Tutoring System
  Dataset},'' \emph{Journal of Machine Learning Research}, 2011.

\bibitem{gunduz2003web}
S.~G{\"u}nd{\"u}z and M.~T. {\"O}zsu, ``{A Web Page Prediction Model Based on
  Clickstream Tree Representation of User Behavior},'' in \emph{KDD}.\hskip 1em
  plus 0.5em minus 0.4em\relax ACM, 2003, pp. 535--540.

\bibitem{speiser2012nested}
M.~Speiser, G.~Antonini, A.~Labbi, and J.~Sutanto, ``{On Nested Palindromes in
  Clickstream Data},'' in \emph{KDD}.\hskip 1em plus 0.5em minus 0.4em\relax
  ACM, 2012, pp. 1460--1468.

\end{thebibliography}

\vspace{-0.4in}
\begin{IEEEbiographynophoto}{Christopher G. Brinton (S'08)}
is a PhD Candidate of Electrical Engineering at Princeton University, Princeton, NJ. His research focus is in big data learning analytics and social learning networks, with significant tech transfer to a learning technology startup. He is also MOOC co-instructor, having co-authored a textbook on social and technological networking. Chris received his Master'€™s Degree in EE from Princeton in May 2013, and his BSEE from The College of New Jersey (valedictorian and summa cum laude) in May 2011.
\vspace{-0.4in}
\end{IEEEbiographynophoto}

\begin{IEEEbiographynophoto}{Swapna Buccapatnam}
is a postdoctoral researcher at the IBM T.J. Watson Research Center, Yorktown Heights, NY, USA. Prior to this, she was a postdoctoral research associate in the Department of Electrical Engineering at Princeton University. She received her Ph.D. in Electrical and Computer Engineering from the Ohio State University in 2014 and her undergraduate degree in Electrical Engineering from the Indian Institute of Technology, Madras, India, in 2008. Her research interests lie in stochastic modeling and analysis, machine learning, data analytics, and optimization.
\vspace{-0.4in}
\end{IEEEbiographynophoto}

\begin{IEEEbiographynophoto}{Mung Chiang (S'00, M'03, SM'08, F'12)}
is the Arthur LeGrand Doty Professor of Electrical Engineering at Princeton University, Princeton, NJ. His research on communication networks received the 2013 Alan T. Waterman Award from the US National Science Foundation, the 2012 Kiyo Tomiyasu Award from IEEE, and various young investigator awards and paper prizes. A TR35 Young Innovator Award recipient, he created the Princeton EDGE Lab in 2009 to bridge the theory-practice divide in networking by spanning from proofs to prototypes, resulting in several technology transfers to industry and two startup companies. He is the Chairman of the Princeton Entrepreneurship Advisory Committee and the Director of the Keller Center for Innovations in Engineering Education. His MOOC in social and technological networks reached about 200,000 students since 2012 and lead to two undergraduate textbooks and he received the 2013 Frederick E. Terman Award from the American Society of Engineering Education. He was named a Guggenheim Fellow in 2014.
\vspace{-0.4in}
\end{IEEEbiographynophoto}

\begin{IEEEbiographynophoto}{H. Vincent Poor (S'72, M'77, SM'82, F'87)}
received the Ph.D. degree in EECS from Princeton University in 1977. From 1977 until 1990, he was on the faculty of the University of Illinois at Urbana-Champaign. Since 1990 he has been on the faculty at Princeton, where he is the Michael Henry Strater University Professor of Electrical Engineering and Dean of the School of Engineering and Applied Science. Dr. Poor's research interests are in the areas of stochastic analysis, statistical signal processing, and information theory, and their applications in wireless networks and related fields such as social networks and smart grid. Among his publications in these areas is the recent book \textit{Mechanisms and Games for Dynamic Spectrum Allocation} (Cambridge University Press, 2014). Dr. Poor is a member of the National Academy of Engineering and the National Academy of Sciences, and a foreign member of Academia Europaea and the Royal Society. He is also a fellow of the American Academy of Arts and Sciences, the Royal Academy of Engineering (U. K), and the Royal Society of Edinburgh. He received the Technical Achievement and Society Awards of the IEEE Signal Processing Society in 2007 and 2011, respectively. Recent recognition of his work includes the 2014 URSI Booker Gold Medal, and honorary doctorates from several universities in Europe and Asia, including an honorary D.Sc. from Aalto University in 2014.
\end{IEEEbiographynophoto}

\end{document}